\documentclass{aa}  
\usepackage[utf8]{inputenc}
\usepackage{graphicx}
\usepackage[english]{babel}
\usepackage{natbib}

\usepackage{txfonts}
\newcommand{\degree}{\ensuremath{^\circ}}

\newcommand{\aleo}{$\alpha$~Leo}
\newcommand{\kms}{\ensuremath{\rm{km\,s^{-1}}}}

\newcommand{\CII}{C~{\sc ii}}

\newcommand{\OI}{O~{\sc i}}
\newcommand{\OVI}{O~{\sc vi}}
\newcommand{\ArI}{Ar~{\sc i}}
\newcommand{\NI}{N~{\sc i}}
\newcommand{\NII}{N~{\sc ii}}
\newcommand{\SiIII}{Si~{\sc iii}}
\newcommand{\SiII}{Si~{\sc ii}}
\newcommand{\HI}{H~{\sc i}}
\newcommand{\DI}{D~{\sc i}}
\newcommand{\SII}{S~{\sc ii}}
\newcommand{\SIII}{S~{\sc iii}}
\newcommand{\MgII}{Mg~{\sc ii}}
\newcommand{\MgI}{Mg~{\sc i}}
\newcommand{\FeII}{Fe~{\sc ii}}
\newcommand{\AlII}{Al~{\sc ii}}
\newcommand{\CIV}{C~{\sc iv}}
\newcommand{\AlIII}{Al~{\sc iii}}
\newcommand{\SiIV}{Si~{\sc iv}}
\newcommand{\NV}{N~{\sc v}}
\newcommand{\CaII}{Ca~{\sc ii}}
\newcommand{\cmc}{cm$^{-3}$}
\newcommand{\cms}{cm$^{-2}$}
\begin{document} 

   \title{The nearby interstellar medium towards \aleo\ --\\
   UV observations and modeling of a warm cloud within hot gas
    }

   \author{Cecile Gry
          \inst{1}
          \and
          Edward B. Jenkins \inst{2}   
 }
   \institute{    Aix Marseille Universit\'e, CNRS, LAM (Laboratoire d'Astrophysique de Marseille) UMR 7326, 13388, Marseille, France\\
              \email{cecile.gry@lam.fr}       
               \and
               Department of Astrophysical Sciences, Princeton University Observatory, Princeton, NJ 08544, USA\\
             \email{ebj@astro.princeton.edu}
                         }

   \date{Received 24 May 2016 ; Accepted 7 September 2016}
   
   \titlerunning{Nearby ISM towards \aleo}
\authorrunning{Cecile Gry \& Edward B. Jenkins }

 
  \abstract {}
 {Our aim is to characterize the conditions in the closest interstellar cloud.}
{We analyze interstellar absorption features in the full UV spectrum of the nearby (d = 24 pc) B8 IVn star alpha Leo (Regulus) obtained at high resolution and high S/N by the HST ASTRAL Treasury program.  We derive column densities for many key atomic species and interpret their partial ionizations.  }
   { The gas in front of alpha Leo exhibits two absorption components, one of which coincides in velocity with the local interstellar cloud (LIC) that surrounds the Sun.  The second, smaller, component is shifted by +5.6 km/s relative to the main component, in agreement with results for other lines of sight in this region of the sky. The excitation of the C II fine-structure levels and the ratio of Mg I to Mg II reveal a temperature T = 6500 (+750,-600)K and electron density n(e) = 0.11 (+0.025,-0.03)\cmc. Our investigation of the ionization balance of all the available species indicates that about 1/3 of the hydrogen atoms are ionized and that metals are significantly depleted onto grains.  We infer that N(H I) = 1.9 (+0.9,-0.6)\,10$^{18}$\cmc, which indicates that this partly neutral gas occupies only 2 to 8 pc (about 13\%) of the space toward the star, with the remaining volume presumably being filled with a hot gas that emits soft X-rays.  We do not detect any absorption features from the highly ionized species that could be produced in an interface between the warm medium and the surrounding hot gas. Finally, the radial velocity of the LIC agrees with that of the Local Leo Cold Cloud, indicating that they may be physically related. 
  }
{}
 \keywords{ISM:  clouds -- ISM: abundances -- (Galaxy:) local interstellar matter
                 -- ISM: individual objects: Local Cloud, LIC, Local Leo Cold Cloud, LLCC-- Ultraviolet: ISM
                 -- Stars: individual: \aleo}
\maketitle
   
%
%
\section{Introduction}
\subsection{General motivation}

Beginning with findings from the {\it Copernicus\/} satellite \citep{Spitzer.Jenkins1975},  studies of ultraviolet absorption features appearing in the spectra of hot, 
rapidly rotating stars have yielded fundamental insights on the compositions and 
physical characters of different phases of the interstellar medium \citep{Savage.Sembach1996}, along with the processes that influence them.   With the exception 
of white dwarf stars and stars with spectral types A and cooler, nearly all of the 
targets are so distant that their sight lines traverse regions with characteristics that 
are substantially different from one another.  As a consequence, the interstellar 
absorption features usually reveal a heterogeneous mix of the imprints of many 
different regions, which can only be separated by chance offsets in radial velocities.  
Nearby stars offer an opportunity to explore a less cluttered situation, but they have 
the drawback that they represent only one or a few regions with very similar 
properties.  Even so, recent investigations of nearby environment have been very 
useful in revealing its dynamics \citep{Redfield.Linsky2008,Gry.Jenkins2014}, 
gas-phase composition \citep{Lehner2003,Redfield.Linsky2004a}, ionization 
state \citep{Jenkins2000}, temperature, and turbulent velocities \citep{Redfield.Linsky2004b}.   In addition, \cite{Oegerle2005}, \cite{Savage.Lehner2006}, and \cite{Barstow2010} have found evidence for the presence of a very hot medium in some 
nearby locations.  A review of many findings on the local medium has been 
presented by \cite{Frisch.Redfield.Slavin2011}. 
 
The earlier UV studies of the local environment had to contend with some 
difficulties.  White dwarf stars were once thought to have featureless spectra that 
could cleanly show interstellar lines, but subsequent investigations have revealed 
unexpectedly high metal abundances in the atmospheres of these narrow-line stars 
caused by radiative levitation  \citep{Barstow2003} and pollution by the infall of 
circumstellar matter \citep{Rafikov.Garmilla2012,Barstow2014}.  These 
atmospheric metal-line features, along with ones arising from circumstellar matter, 
can create serious confusion when attempting to discern interstellar features 
\citep{Lallement2011}, but in a small percentage of cases the interstellar 
absorptions can be separated from the photospheric or circumstellar contributions 
\citep{Barstow2010}.  For cool stars, a different problem emerges.  Here, interstellar 
absorption features must be viewed on top of chromospheric emission features, a 
problem which forces one to have a good understanding of the shapes of 
the underlying emission profiles.

O and B type stars whose stellar features are broadened by rotation represent ideal 
targets for interstellar absorption line studies.  This paper focuses on the UV 
spectrum of one such star,  \aleo\ (Regulus), a bright (V~=~1.40) B8\,IVn star
that was recently observed in a
230-orbit HST Cycle 21 Treasury Program (program ID = 13346, T.~R.~Ayres, PI) 
called the Advanced Spectral Library II: Hot Stars (ASTRAL).  This observing 
program produced atlases of high resolution, complete UV spectra of 21 diverse 
early-type stars.  Our target $\alpha$~Leo at a distance of 24\,pc is the nearest
B-type main sequence or giant star in the sky.  Its strong brightness in the 
ultraviolet and its high projected rotational velocity ($v\sin i=353\,\kms$) make it an 
ideal target for investigating the local medium.  As we discuss in later sections of 
this paper, the spectrum of this star reveals important details on the density and 
degree of ionization of hydrogen atoms, the gas-phase abundances of certain 
elements, the temperature of the gas, along with the filling factor along the sight line 
for constituents that we can detect.  Finally, at the Galactic coordinates 
$\ell=226.4\degree$, $b=+48.9\degree$, \aleo\ samples a portion of the sky that 
has not been well sampled at close distances \citep{Gry.Jenkins2014,Malamut2014}.

\subsection{Our local environment}

In broadest terms, the Sun is located in a specially rarefied region of the Galaxy. It is 
situated in a small, diffuse ($n_\mathrm{H~I} = 0.05-0.3$ cm$^{-3}$), warm 
medium, which itself is embedded in an irregularly shaped cavity of about 100~pc 
radius \citep{Welsh2010,Lallement2014}. This cavity is called the Local 
Bubble. It is almost devoid of neutral gas and is probably filled mostly with a hot 
($T\sim 10^6$ to $10^7$\,K) tenuous, collisionally ionized gas that emits soft
X-rays \citep{Williamson1974,McCammon.Sanders1990,Snowden1997,Snowden2014}. 

It is generally recognized that the Local Interstellar Cloud (LIC) that surrounds our 
heliospheric environment is partly ionized to a level $n_e/n_{\rm H} \sim 0.5$ by 
the ambient EUV and X-ray radiation field that arises from stars \citep{Vallerga1998} and 
the surrounding hot gas \citep{Slavin.Frisch2008}.  In accord with previous findings 
by \cite{Redfield.Linsky2004b}, we will show that the temperature of the gas is 
$T\sim 7000\,$K, which is one of the stable phases that arises from the bifurcation due  
to the Field (1965) thermal instability \citep{Wolfire2003}.  The magnetic field at 
distances greater than 1000\,AU from the Sun has a strength of about $3\mu$G and 
is directed toward $\ell=26.1\degree$, $b=49.5\degree$, according to an 
interpretation of results from the {\it Interstellar Boundary Explorer\/} (IBEX) by 
\cite{Zirnstein2016}.  
The magnetic direction thus makes an angle of about 79 \degree\  with respect to the direction toward \aleo. 

The distance to the boundary between the LIC and the surrounding hot medium is 
not well determined, but it is probably of order 10\,pc \citep{Frisch.Redfield.Slavin2011,Gry.Jenkins2014}.  It is therefore very likely that the sight line to \aleo\ penetrates this 
boundary.  In principle, UV spectroscopic data should help us to understand the 
nature of this boundary: is it a conduction front where evaporation or condensation 
of warm gas is occurring \citep{Cowie.McKee1977,McKee.Cowie1977,Ballet1986,Slavin1989,Borkowski1990,Dalton.Balbus1993}?  
Alternatively, could it be a turbulent mixing layer (TML), where, as the name 
implies, the existence of any shear in velocity between the phases creates 
instabilities and mechanically induced chaotic interactions \citep{Begelman.Fabian1990,Slavin1993,Kwak.Shelton2010}?  Observers have attempted to 
identify these processes chiefly by analyzing interstellar absorption features of ions 
that are most abundant at intermediate temperatures, such as Si~IV, C~IV, N~V and 
O~VI, and then comparing their column density ratios with the theoretical 
predictions \citep{Spitzer1996,Sembach1997,Zsargo2003,Indebetouw.Shull2004a,Indebetouw.Shull2004b,Lehner2011,Wakker2012}.  
Such studies have been conducted over very long sight lines, where multiple 
interfaces may be found.  Also, the signatures from interfaces possibly could be mixed with contributions from radiatively 
cooling gases. Papers that report these results give us information on the relative 
importance of different cases, but they tell us nothing about what happens within 
any single interface.

Inside the Local Bubble, there are a few isolated, dense clouds \citep{Magnani1985,Begum2010}. One such cloud attracted the attention of \cite{Verschuur1969} and 
was studied further by \cite{Verschuur.Knapp1971} because it had an unusually low 
21-cm spin temperature and a low velocity dispersion. This cloud covering 
22~$\deg^2$ in the Leo constellation and later estimated to be at a distance 
between 11 and 40 pc away from us was investigated using optical absorption 
features by \cite{Meyer2006} and in further detail by \cite{Peek2011}, who 
named the cloud the Local Leo Cold Cloud (LLCC).  The upper limit to its distance of 
40 pc was established by \citep{Meyer2006}, who detected narrow Na I features 
toward two stars beyond the LLCC located at distances slightly over 40 pc.  From 
HST STIS spectroscopy of stars behind this cloud, \cite{Meyer2012} analyzed the 
excitation of the fine-structure levels of C~I and came to the remarkable conclusion 
that this cloud had an internal thermal pressure $p/k \approx 60,000\,{\rm cm}^{-
3}\,$ K, which is considerably higher than estimates of $p/k \leq 10,000\,{\rm 
cm}^{-3}\,$K for the low density material inside the Local Bubble \citep{Jenkins2002,Jenkins2009,Frisch.Redfield.Slavin2011,Snowden2014}. From its extraordinarily 
high thermal pressure and low 21-cm spin temperatures [13 to 22 K \citep{Heiles.Troland2003}], this cloud presents an anomaly in the very diffuse context of the 
Local Bubble. Another interesting feature of this cloud is that it appears to coincide 
with a long string of other dense clouds that stretches across 80\degree\ in the sky 
\citep{Haud2010}.

Our target \aleo\ is separated in projection by only about 4\degree\ from the LLCC.  In 
fact, \cite{Peek2011} claimed that the very weak Ca II absorption feature in the 
spectrum of \aleo\ arises from the outermost portions of the LLCC, and therefore 
proposed that the LLCC distance upper limit be determined by the distance to the 
star \aleo. However we argue later that this component is likely to be produced by 
the local interstellar cloud surrounding the Sun.


\section{Observations and spectral analysis\label{sec:observations}}
\subsection{HST-STIS data}
Our spectrum was recorded with the E140H and E230H echelle modes at a spectral resolution $\lambda/\Delta \lambda\approx 114,000$ over the wavelength interval 1164 to 3045\AA. Many of the observations in the ASTRAL program were performed with the neutral density (ND) 0.2x0.05 arc-sec entrance slit because the stars were very bright. This had to be done for \aleo. This slit is narrower than the standard 0.2x0.09 arc-sec slit.  Initially, we used the CALSTIS reduction package to process the data in a mode that recognized intensities within half pixels of the MAMA detector in order to retrieve the full wavelength resolution of the original data. 
However, we found that the very modest gain in resolution over the standard full pixel sampling was more than offset by the compromises from the greater uncertainties in the wavelength calibration and the zero-flux level determinations. We therefore decided to use simply the normal spectral sampling provided by the Mikulski Archive for Space Telescopes (MAST), which had better calibration files and improved control over the systematic errors.

With a few exceptions, we did not perform multi-exposure co-additions. For multiple exposures or wavelengths where there were order overlaps, we retained each individual spectrum segment for analysis so that we could avoid the misleading effects that arise from the interpolated resamplings, which introduce correlated noise in the spectra. Our independent treatments of duplicate spectral coverages secured separate error estimates and allowed us to obtain reliable estimates of the $\chi^2$ values which we could then combine in the ultimate minimization process for the line fitting. This also maintains the full resolution, since there is no resolution degradation introduced by interpolation of the wavelength grids or slight wavelength shifts over the different exposures. We found that for the strongest saturated lines we could sense slight errors in the derivations for the zero-flux levels.  Thus, by keeping the different exposures separate we could correct for these errors individually and avoid systematic errors that might be hidden in the co-added spectra. The only cases where we had to use co-addition was for the Lyman-$\alpha$ profile discussed in Section~\ref{sec:HI}, where the noise is so high that exposures cannot be reliably interpreted individually, 
as well as to derive column density upper limits for the undetected lines \CIV, \NV, \AlIII, \SiIII, \SiIV\ and \SIII.

\subsection{ Copernicus data \label{sec:copernicus}}

Our target star \aleo\ was one of the first set of stars observed in 1973 with the {\it Copernicus\/} satellite \citep{Rogerson1973a}.  The results from these initial observations were reported by \cite{Rogerson1973b} in their study of what they called the ``intercloud" medium. While working on the STIS data, we realized that more extensive {\it Copernicus\/} spectra were acquired in 1977, which have not yet been published. They included the \OI\ line  at 1039 \AA, which is weaker and hence less saturated than the line at 1302 \AA\ that we obtained with STIS.  This weaker line was therefore quite beneficial in constraining the column density of O~I.  In addition, the {\it Copernicus\/} spectra covered the \NII\ 1084\,\AA\ line and the two \ArI\ lines at 1048 and 1066 \AA, all of which are outside the wavelength coverage of STIS. 
At the time of the observations taken in 1977,  the stray light was blocked and the only sources of background were charged particles and scattered light from the grating, making the background easier to correct than with the earlier published data.

We downloaded the co-added spectra from the MAST archive and used them to supplement our STIS data.
We estimated background corrections for the {\it Copernicus\/} spectra through the use of Eq.~2 of \cite{Bohlin1975}, which expresses the scattered light level in the U1 tube at a wavelength $\lambda$ in counts according to the relation
\begin{equation}
{\rm G}_1(\lambda) = 0.02 [{\rm U}_1(1200 \pm 200)] + 0.067 [{\rm U}_1(\lambda) \pm 12]~,
\end{equation}
where the brackets indicate an average perceived count level over the wavelength range specified. According to \cite{Bohlin1975} the worst-case error is 2\% of the continuum. 
For the absorption feature of N~II depicted in Figure~\ref{fig:fits} the background is thus 56.8 $\pm$ 7.2. It is 24.6 $\pm$ 4 for \OI\ 1039\AA, 39.7 $\pm$ 8.8 for \OI\ 1302\AA, 49.10 $\pm$ 18 for \ArI\ 1066 \AA\ and 36.4 $\pm$ 9 for \ArI\ 1048\AA. The uncertainties in the background levels are the principal sources of error for the column densities that we derived. (The panels of Fig.~\ref{fig:fits} show the corrected count rates). 

%

\subsection{Spectral analysis: multi-element, multi-component profile fitting\label{sub:line-fitting}}
To derive the characteristics of the interstellar components,
i.e. column density $N $, velocity $ v $ and broadening parameter $b $, 
we compared the observed line profiles with theoretical ones that represent the convolution of a Voigt profile with the instrumental line spread function (LSF).
This comparison was performed with the use of the profile fitting software
{}``Owens.f'' developed in the 1990's by Martin Lemoine and the French FUSE team.
For all species, we adopted the $f$-values listed by \cite{Morton2003}.

\begin{figure*}[!htbp]
\begin{center}
\includegraphics[width=1.95\columnwidth]{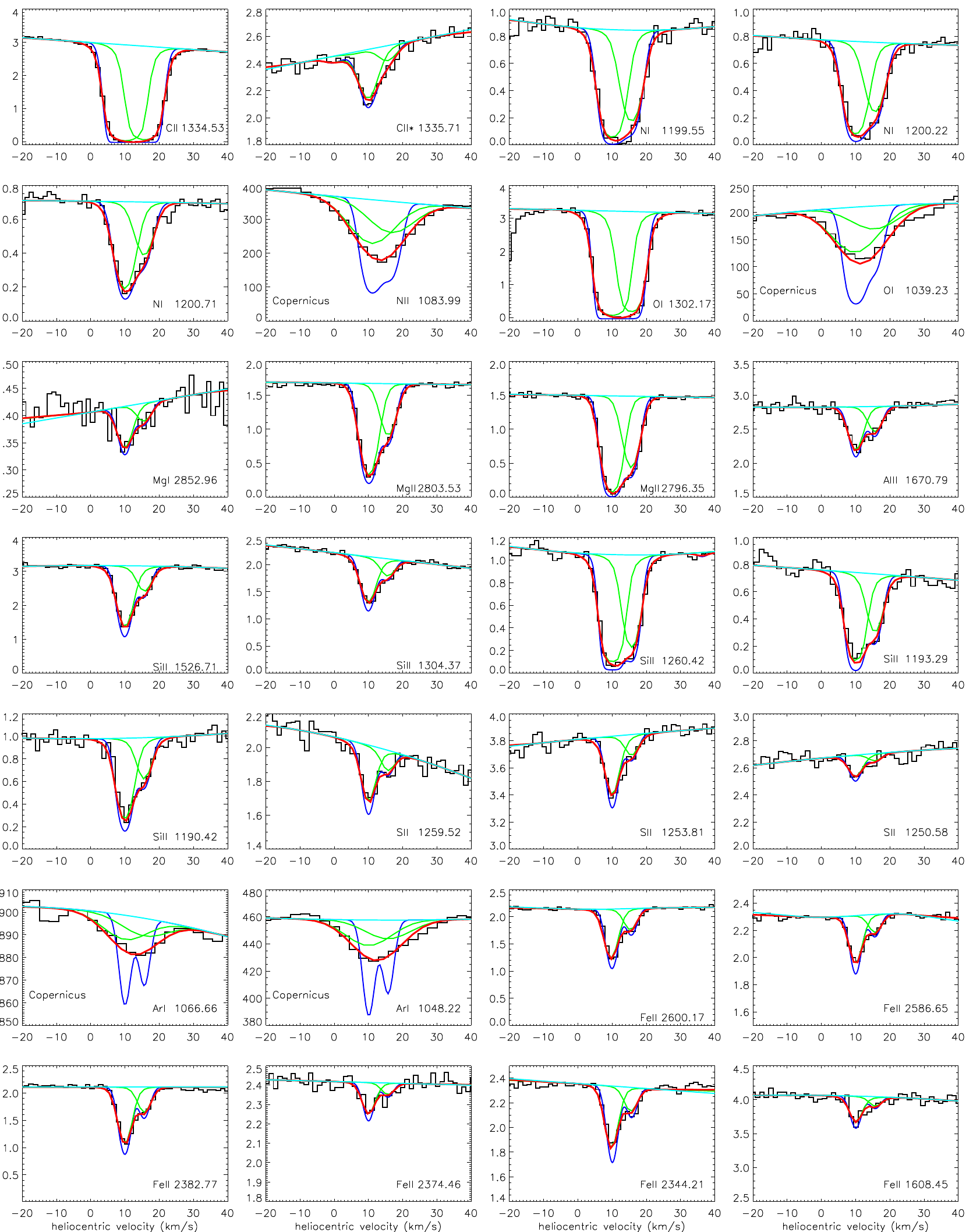} 
\end{center}
\caption{Absorption line spectra  in the line of sight to \aleo.
Black histogram-style curves represent the observations,
red solid lines are the best fits, green lines show the contributions from the two individual components that we could identify, and blue lines represent our reconstructions of the profiles before convolution
with the instrumental LSF, i.e. the intrinsic interstellar profile. Stellar continua are shown in cyan.
All fits have been performed simultaneously with the model for the line of
sight with 2 components having same velocity, same temperature, and same turbulent broadening for
all species. Copernicus spectra are labeled ``Copernicus'', their flux scale is expressed in counts per 14s ; the remaining spectra are from HST/STIS and are indicated in the units $10^{-9}\,{\rm erg \,{\rm cm}^{-2}\, sec^{-1}\,\AA^{-1}}$.\label{fig:fits}}
\end{figure*}

The software allows one to simultaneously fit several lines of the same element,
as well as lines from different elements. It also provides for the existence of 
several velocity components that can be fitted simultaneously yielding the characteristics of individual velocity components even when their profiles are blended together. %
The software therefore derives a global and consistent
solution for all species with a common absorption velocity and common
physical conditions, implying consistent broadening parameters.
The fitting software breaks the line-broadening parameter ($b$-value, 
 i.e. ${\sqrt(2) \times \sigma(v)}$ projected along the sight line)
into thermal broadening, which depends on the element mass, and
non-thermal broadening (turbulence), which is the same for all elements
in any given component.  
Therefore, fitting lines from elements of different masses simultaneously in principle enables the 
simultaneous measurement of temperature, turbulent velocity, mean velocity, and element column densities.
All detected spectral lines used in this analysis  are
 presented in Figure~\ref{fig:fits} together with their fits.

 The spectra have not been normalized to the stellar continuum; instead the stellar profile is included in the fit as $n+1$ free parameters for an $n$-degree polynomial (usually a straight line).
 
The LSFs applying to the Echelle gratings E140H and E230H are tabulated in the STIS Instrument
Handbook for the two standard high-resolution slits 0.2x0.09 arc-sec and 0.1x0.03 arc-sec. They are represented by a sum of broad and narrow components whose FWHMs we derive in each case by performing a double-Gaussian fit to the tabulated LSFs. Since \aleo\ has been observed with the non-conventional slit 0.05ND for which the LSF has not been tabulated, and since its width is intermediate between the two standard slits, we make the assumption (recommended by STScI specialists) that the FWHMs for this slit can be interpolated from those resulting from the double-Gaussian fit to the two tabulated LSFs. As a result, we adopted for the low amplitude broad component a FWHM of 6 pixels and 5.11 pixels respectively for E140H and E230H data, and for the taller narrow component an FWHM that varies slightly with wavelength: 0.91 pixels at 1200 \AA, 0.89 pixels at 1500 \AA\ (E140H), and 1.50 pixels at 2400 \AA\ (E230H).  In essence, the spectra are undersampled according to the Nyquist criterion and thus are subject to aliasing.  Our experiences with the half-pixel sampling indicated that this is not a serious problem.
 For {\it Copernicus} data, we have adopted a Gaussian LSF with a FWHM of 0.062 at 1302\AA\ and 0.051 for the other lines, corresponding to a resolution of 20\,000 everywhere.\\  
According to the STIS Instrument Handbook, the wavelength
accuracy across exposures is 0.2-0.5 pixels. We do observe slight velocity variations in the spectra. Therefore, when we fit the spectra we allow for a free velocity shift between the different wavelength windows that cover different spectral lines or for identical spectral lines in different orders or exposures. The resulting velocity shifts have a dispersion close to 0.5\,\kms, so they are indeed generally lower than 0.5 pixel or 0.66\,\kms, except in rare cases such as near 1304\AA\ in order 12 of both E140H-1271\AA\ exposures,  where the measured wavelength shift is $\sim$ +0.9\,\kms\ at the position of the O~I* and Si~II lines.

We can check on the heliocentric velocities of the interstellar components by measuring the observed velocities of the telluric absorption features of O~I that originate from oxygen atoms in the Earth's upper atmosphere.  Such features should be offset from the heliocentric velocities by a correction listed in the V\_HELIO keyword given in the data headers.
On our reference O~I spectrum (not allowed to shift in wavelength during the fit process) for which the keyword V\_HELIO = 22.23 \kms, we measure a telluric velocity of $-21.21\,\kms$. From this offset we derive an absolute offset $\Delta$V=1.02 \kms\ that we subsequently subtract from the fit results to ascertain the absolute velocities of the interstellar components.\footnote{We are unable to correct for rotation of the Earth's atmosphere. For the declination of \aleo\ (12$^o$) the telluric features can experience a shift as large as 0.45\,\kms\ when viewed at large zenith angles.}

The  {\it Copernicus} data have a much larger velocity uncertainty caused by changes in the spectrometer temperatures (whose values are not available in the archive). The fits to the {\it Copernicus} lines therefore also allow a free velocity shift that can be as high as 8\,\kms. For this reason, along with the lack of resolution, we had to declare that the allotments of column densities between the two principal velocity components are uncertain for lines that are only observed with {\it Copernicus}.
 
    \begin{table*}[!htbp]
      \caption[]{Results of the profile fitting for the two components in the \aleo\ sight line. The intervals indicate 1-$\sigma$ uncertainties.

         \label{tab:results}}
         \begin{tabular}{lcccc}
            \hline 
            &&&&\\
  Component &   1 (LIC) &    2 & total line of sight & $N$(LIC)/$N_{\rm total}$\\
               \hline 
                           &&&&\\

V$_{\rm heliocentric}$ (\kms)&	8.8	$\pm$ 0.2	&	14.4$\pm$	0.1  &&\\
V$_{\rm LSR}$ (\kms)&	2.7	$\pm$ 0.2	&	8.3$\pm$	0.1  &&\\
$T$ (K)	 &	6000 $\pm$600  	&	5990 $\pm$700 &&\\
$b_{\rm turb}$	(\kms) &	1.78 $\pm$0.10	&1.85 $\pm$0.19 &&\\
            \hline 
         &&\\
$N$(\CII)& 	     4.5 $^{+1.5}_{-1.0}\,10^{14}$&	2.2$^{+1.4}_{-0.4}\,10^{14}$&6.7$^{+1.4}_{-0.8}\,10^{14}$&0.67$^{+0.30}_{-0.29}$\\[2pt]
$N$(\CII*)&	2.9$\pm$0.2 $10^{12}$&	6.6 $^{+1.6}_{-2.0}\,$$10^{11}$&3.54$\pm$0.16 $10^{12}$&0.82$\pm0.09$\\[2pt]
$N$(\CIV)&	$< 5\,10^{11}$&$<$5 \,$10^{11}$&&\\[0.pt]
$N$(\CIV)$^a$(T=200\,000)&&&	$<$2 \,$10^{12}$&\\[3pt] 
$N$(\NI)&  	7.2 $\pm$0.2 $10^{13	}$&	3.3 $\pm$0.2 $10^{13}$&1.05$\pm$0.03 $10^{14}$&0.69 $\pm 0.04$\\
$N$(\NII)& &&7.0$^{+4.2}_{-1.0}\,10^{13}$&\\[1.5pt]
$N$(\NV)&	$<$ 6.8 $10^{11}$&	$<$ 6.8 $10^{11}$&\\[0pt]
$N$(\NV)$^a$(T=200\,000)&&&	$<$ 2.5 $10^{12}$&\\[3pt]
$N$(\OI)&	     9.5 $^{+1.4}_{-1.2}\,10^{14}$& 2.0 $\pm$0.2 $10^{14}$&1.15$^{+0.15}_{-0.13}\,10^{15}$&0.83$^{+0.22}_{-0.21}$\\[1pt]
$N$(\OVI)$^b$&	     & &$<$ 3.1\,$10^{12}$&\\[3pt]
$N$(\MgI)& 	2.3 $\pm$0.7 $10^{10}$&	1.0 $\pm$0.7 $10^{10}\,^{\it c}$&3.3$^{+0.6}_{-0.8}\,10^{10}$&0.70$^{+0.38}_{-0.34}$\\[1pt]
$N$(\MgII)&	     4.9$\pm$0.2 $10^{12}$&	1.64 $\pm$0.06 $10^{12}$&6.5$\pm 0.2 \,10^{12}$&0.75$\pm 0.05$\\[3pt]
$N$(\AlII)&	1.85 $\pm0.17\,10^{11}$&	1.00 $\pm0.17\, 10^{11}$&	2.84 $\pm 0.11\,10^{11}$&		0.65 $\pm0.09$\\[1pt]
$N$(\AlIII)&	$<$ 6.8$10^{10}$&	$<$ 6.8 $10^{10}$&\\[3pt]
$N$(\SiII)&     1.00 $\pm$0.06 $10^{13}$&	3.02 $\pm$0.30 $10^{12}$&1.30$\pm$0.06 $10^{13}$&0.77$\pm 0.08$\\[1pt]
$N$(\SiIII)& $<$ 8.0\,$10^{11}$&$<$ 8.0\,$10^{11}$&&\\[1pt]
$N$(\SiIV)&	$<$ 1.6 $10^{11}$&	$<$ 1.6 $10^{11}$&\\[1pt]
$N$(\SiIV)$^a$(T=200\,000)&&&	$<$ 4 $10^{11}$&\\[3pt]
$N$(\SII)&	     1.94$\pm$0.21 $10^{13}$&	4.9 $\pm$2.7 $10^{12}\,^{\it d}$&2.44$\pm$0.24 $10^{13}$&0.80$\pm 0.16$\\ [1pt]
$N$(\SIII)&	$<$ 8.8 $10^{12}$&	$<$ 8.8 $10^{12}$&\\[3pt]
$N$(\ArI)&&&1.86$\pm 0.07 \,10^{12}$&\\[3pt]
$N$(\CaII)$^e$&&&5 $10^{9}$&\\[3pt]
$N$(\FeII)&    1.82$\pm$0.03 $10^{12}$&	6.3 $\pm$0.3 $10^{11}$&2.45 $\pm$0.03 $10^{12}$&0.74$\pm 0.02$\\[3pt]
  \hline
\multicolumn{3}{l}{$N$(\HI)$^f$ (apparent column density for the 2-component model)}&    {6.6    $\pm$1.0  $10^{18}$}$^f$&\\[1pt]
  \hline

\multicolumn{5}{l}{$^{a}$ {\footnotesize estimated for T$\simeq$ 200\,000, in the case it originates in an interface between the cloud and surrounding hot gas}}\\
\multicolumn{5}{l}{$^{b}$ {\footnotesize from \cite{Rogerson1973b}}}\\
\multicolumn{5}{l}{$^{c}$ {\footnotesize undetected at 2-$\sigma$: $<$2.5\,$10^{10}$}}\\
\multicolumn{5}{l}{$^{d}$ {\footnotesize  undetected at 2-$\sigma$: $<$1\,$10^{13}$}}\\
\multicolumn{5}{l}{$^{e}$ {\footnotesize from \cite{Frisch2002}}}\\
\multicolumn{5}{l}{$^{f}$ {\footnotesize  Based on the Lyman-$\alpha$ absorption profile, which we believe overstates $N$(H~I).  See Section~\ref{sec:HI}}}\\
\end{tabular}
\end{table*} 
    
\subsection{Results from profile fitting \label{sec:results}}
The quality of the fits, and  the 
error bars on column densities, have been computed using the
$\Delta\chi^{2}$ method described e.g. in \cite{Hebrard2002}. We performed
several fits by fixing the column density of a given element ${\rm X}$
to a different value in each fit. In all of these fits, all other
parameters are set free and we compute the best $\chi^{2}$ for each value of
column density. Then plots of $\Delta\chi^{2}$ versus $N(X)$ yield the
1, 2, or 3-$\sigma$ ranges for $N({\rm X})$. 

The fits to highly saturated features, like C~II and to a lesser extent O~I, are constrained by 
 the fact that they are performed together with the less saturated lines and the requirement that all share  a  consistent  solution. 
Because we make the assumption that all species share the same turbulence and temperature, we expect that the values of these quantities for the strong lines are relatively well constrained by unsaturated lines.  We use this principle to determine the column densities of species exhibiting only strong lines, ones that would otherwise be impossible to measure with any reasonable accuracy.

The column densities arising from the fits are listed in Table~\ref{tab:results}. The quoted uncertainty intervals include the effects of  wavelength calibration uncertainties, as well as errors in the stellar continuum or detector zero-level placements, since variations of these three factors are permitted in the fit. The uncertainties however do not take into account uncertainty on our knowledge of the LSF, but we have checked that at the level of accuracy to which the LSF is known, the effects of such uncertainties are negligible. 

We have used co-added spectra to derive upper limits for the column densities of undetected species in the following lines: \CIV\,1548 \&\ 1550\AA, \NV\,1238 \&\ 1242\AA, \AlIII\,1854 \&\ 1862\AA, \SiIII\,1206\AA, \SiIV\,1393 \&\ 1402\AA\ and \SIII\,1190\AA. 
At a most fundamental level, uncertainties caused by photon counting noise can alter the shape of an absorption feature, and a means for estimating an error from this process for an equivalent width
has been outlined by Jenkins et al. (1973).  Adding to this uncertainty is the error in establishing
a continuum level.  
For the special cases associated with weak, broad features that might be expected for
highly ionized species, these two uncertainties are overshadowed by a low frequency, fixed-pattern
noise arising from sensitivity variations in the STIS image sensor.  These perturbations create random
disturbances in the apparent fluxes that could obliterate (or masquerade as) real absorption features.
For this reason, we defined upper limits for the high ions  in terms of features that
would be strong enough to overpower in a convincing way these fluctuations.  The characteristic amplitudes
of these disturbances were gauged by examining the fluxes 
in wavelength regions somewhat removed from the expected locations of the interstellar lines.
Our measurements applied to wavelength intervals that were appropriate for our assumed b-values produced by
Doppler broadening at a temperature $T$ combined with turbulent broadening characterized by $v_{\rm turb}$.
On the premise that the highly ionized species such as \CIV, \SiIV\ and \NV\ might arise from an interface between the cloud and the surrounding hot gas instead of photoionization within the cloud itself, we also quote the upper limits inferred for a $b$-value corresponding to $T\sim 200\,000\,$K.
For  \OVI\ \cite{Rogerson1973b} quoted an upper limit of log N(\OVI)$<$ 12.49.

 In the \aleo\ spectrum the \HI\ interstellar absorption occurs at the bottom of a strong stellar line, where the flux is between 15 and 100 times lower than elsewhere in the spectrum. Therefore the relative noise level precludes us from obtaining any information from the neighboring \DI\ line. 

The damped Lyman-$\alpha$ line should in principle allow a precise determination of the total \HI\ column density, but without conveying information on the velocity distribution of the gas. 
We recognize however that the \HI\ column density results can be distorted by the presence of very small amounts of high-velocity or high-temperature gas, either in the immediate vicinity of the star or further away, especially when the column density is low 
as is the case here.
When fitted with the two-component model together with the other elements, the resulting \HI\ column density  in the full line of sight is {6.6    $\pm$1.0  $10^{18}$}, but the reality of this  ``apparent'' \HI\ column density, as stated in Table~\ref{tab:results},  will be questioned in Section~\ref{sec:HI}.

Generally, our column densities agree with those determined by \cite{Rogerson1973b} to within our respective errors.  Two exceptions to this are the measurements of Ar~I and Fe~II.  Our determination for $N$(Ar~I) is 0.39~dex lower than that reported by \cite{Rogerson1973b}, and our value for $N$(Fe~II) is 0.40~dex higher than the earlier measurement.\footnote{The {\it Copernicus\/} observations of the Fe~II lines were carried out using either the V1 or V2 detectors, both of which had seriously compromised photometric precision caused by extraordinarily large backgrounds arising from charged particle radiation interacting with the photomultiplier entrance windows \cite{Rogerson1973a}.}  A mild disagreement with the earlier determinations of $N$(Mg~II) can be attributed to probable errors in the adopted $f$-values for the doublet near 1240\,\AA.
\section{Kinematics and distribution of the gas toward \aleo\ \label{sec:kinematics}}
As depicted in Figure~\ref{fig:fits}, two velocity components are detected in the line of sight towards \aleo. They are separated in velocity by 5.6\,\kms. 

Recently \cite{Gry.Jenkins2014} (hereafter GJ14) showed that the whole set of kinematical data from UV spectroscopy of nearby stars is compatible with the picture of a single monolithic cloud called the Local Interstellar Cloud\footnote{A note on nomenclature: In the present article, we retain the designation Local Interstellar Cloud (LIC) for the medium surrounding the Sun, which is a convention followed by many authors in the past.  However, we envision that it refers to a single, coherent medium defined by GJ14 that accounts for most of the matter out to a distance of many tens of pc.  This term is not to be confused with a narrower definition of the LIC proposed by \cite{Redfield.Linsky2008}, who declared that the nearby gas is broken into many distinct clouds, one of which they called the LIC.} that surrounds the Sun in all directions.  This is in contrast to a previous description \citep{Lallement1995, Redfield.Linsky2008, Frisch.Redfield.Slavin2011} where we are located at the edge of a more confined medium that does not extend very far past the Sun in the  direction of the Galactic Center and that is part of a group of cloudlets moving with slightly different velocities. 
In the direction of \aleo\ a projection of the GJ14 model for the LIC\footnote{This model proposes a mean velocity vector of magnitude V~=~25.53\,\kms\ pointing toward the Galactic coordinates $ l=185.84 \degree ; b= -12.79 \degree$, which is nearly identical to the flow that is impacting the heliosphere \citep{Moebius2004,McComas2012}.The model also includes deviations from the mean vector to account for the cloud deceleration and deformation.
} predicts a velocity of  +9.2 $\pm$1.2\,\kms.
Therefore our Component~1, which deviates  by only 0.4 \kms\ from the expectation or the model, is identified as the LIC. 
Component 2 has a positive velocity shift relative to the LIC. This is in agreement with the previous observations in this region of the sky, as shown in Figure 12 of GJ14, reproduced with a few additions in Figure~\ref{fig:second} that appears 
later in Section~\ref{sec:second}. GJ14 have shown that  when a line of sight included an extra absorption component other than the LIC, in half of the sky this component had a velocity shift of around $-7\,\kms$ relative to the LIC, which suggests an implosive motion of gas progressing toward the interior of the cloud. In the second half of the sky, where \aleo\ is located, the extra components are shifted positively relative to the Local Cloud, which is consistent with gas moving away from the cloud interior. This is the case in the direction of \aleo, as is the case for all surrounding local sight-lines from the sample of \cite{Redfield.Linsky2002} included in the study of GJ14. 

We conclude that the two interstellar components detected towards \aleo\ are entirely consistent with expectations for the local interstellar  kinematics.
We will discuss later in Section~\ref{sec:LLCC} the  remarkable kinematical agreements of this gas with the Leo Local Cold Cloud (LLCC) observed a short distance away in the sky from \aleo.

The last column in Table~\ref{tab:results} indicates the proportion of warm gas in the line of sight that is at the LIC velocity. Over the 10 elements where the column densities have been measured separately for both components, the average value $<N$(LIC)/$N_{\rm total}>$ = 0.74 with a dispersion of 0.06. Therefore about 3/4 of the warm gas in the line of sight is in the LIC, and 1/4 in the second velocity component. 
We note furthermore from the low dispersion of the ratio $N$(LIC)/$N_{\rm total}$  that the column density distribution among both components  does not vary significantly from one element to the other. 
This indicates that the conditions are likely to be  similar in both components.
\section{Temperatures and electron densities from observations of magnesium 
and carbon}\label{sec:Mg_C}

\subsection{Ionization equilibrium of Mg}\label{sec:Mg_ioniz_equilib}

We can draw upon our observations of the ratio of neutral and singly ionized 
magnesium to solve for the electron density $n(e)$ using the equation,
\begin{equation}\label{equilib1}
 [\Gamma({\rm Mg}^0)+C({\rm Mg}^+,T)n({\rm H}^+)]n({\rm Mg}^0) = 
\alpha({\rm Mg}^+,T)n(e) n({\rm Mg}^+)~,
\end{equation}
where 
\begin{equation}\label{equilib2}
\Gamma({\rm Mg}^0) =\int_{\rm IP(Mg^0)}^{13.6\,{\rm eV}}\sigma({\rm 
Mg}^0,E)I(E)_{\rm ISRF}dE~,
\end{equation}
$\sigma({\rm Mg}^0,E)$ is the photoionization cross section of neutral Mg for 
photon energies $E$ above the ionization potential ${\rm IP(Mg^0)=7.65\,eV}$, and 
$I(E)_{\rm ISRF}$ is the strength of the interstellar radiation field in the local 
neighborhood (expressed in ${\rm photons~cm}^{-2}{\rm s}^{-1}{\rm eV}^{-1}$ 
integrated over $4\pi\,$str).  The quantity $C({\rm Mg}^+,T)$ is the rate constant for 
the charge exchange reaction ${\rm Mg}^0 + {\rm H}^+ \rightarrow {\rm Mg}^+ + 
{\rm H}^0$, and $\alpha({\rm Mg}^+,T)$ is the sum of the radiative and 
dielectronic recombination rates for free electrons and ionized Mg at a temperature 
$T$.  The equilibrium expressed in Eq.~\ref{equilib1} should hold as long as the 
physical conditions do not change more rapidly than the $e$-folding time 
$[\alpha({\rm Mg}^+,T)n(e)+\Gamma({\rm Mg}^0)+C({\rm Mg}^+,T)n({\rm 
H}^+)]^{-1}\approx 750\,$yr for changes in ratio of ${\rm Mg}^0$ to ${\rm 
Mg}^+$ (at $T=6000\,$K).  In principle, a more comprehensive treatment should 
also include the effects of Mg ions being neutralized when they collide with dust 
grains \citep{Weingartner.Draine2001}, but this effect is at most about 2\% of 
recombination rate with free electrons under typical conditions that we consider: 
$T\approx 6000\,$K, $n(e)\approx 0.1\,{\rm cm}^{-3}$ and $n({\rm H}^0+{\rm 
H}^+)\approx 0.3\,{\rm cm}^{-3}$ \cite{Gry.Jenkins2014}.  

\subsubsection{Far ultraviolet interstellar radiation 
Field}\label{sec:FUV_Rad_Field}

\begin{figure}
\includegraphics[width=1.0\columnwidth]{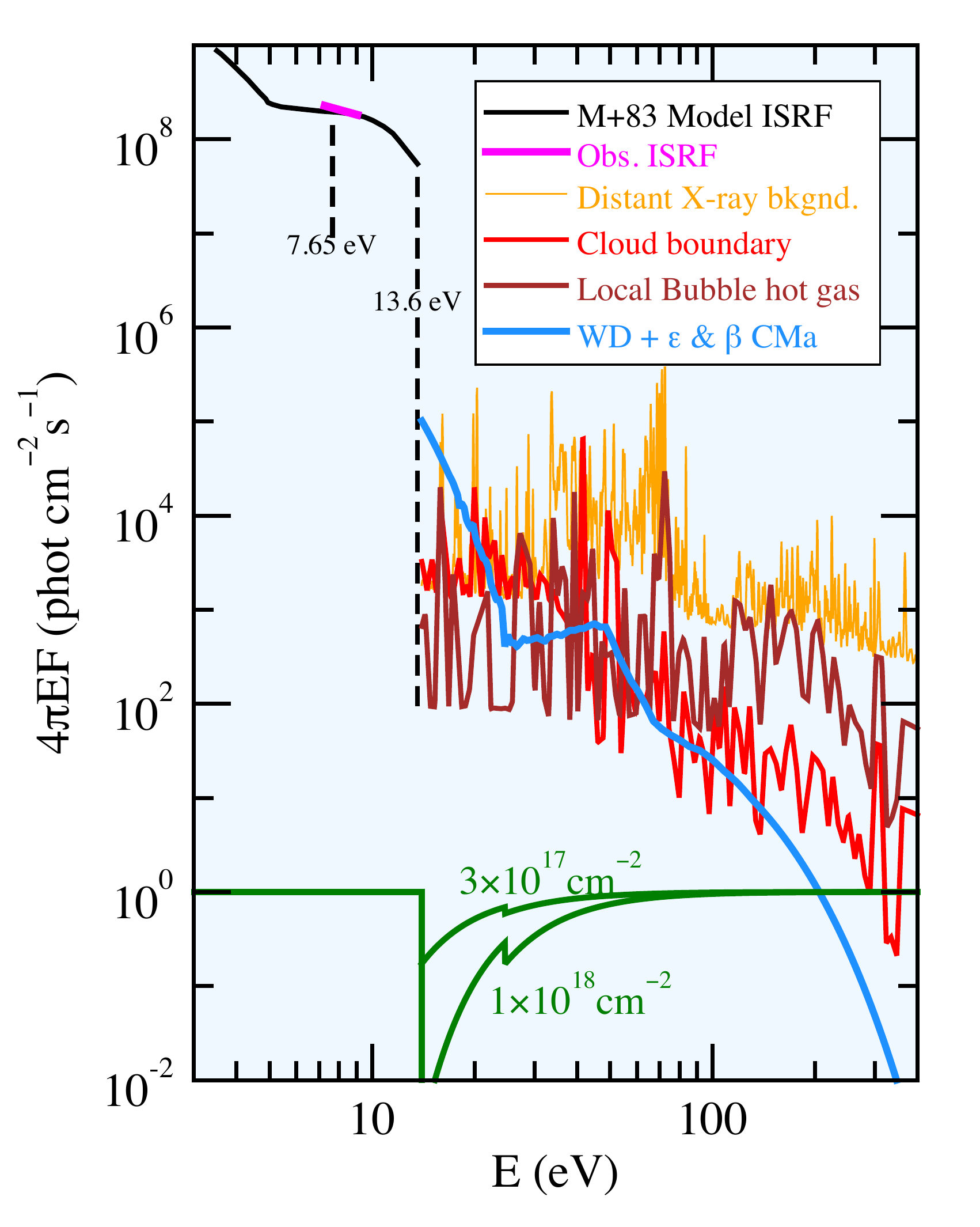}
\caption{The product of energy and the fluxes of ionizing photons $E_{\rm 
eV}I(E)_{\rm ISRF}$ as a function of energy $E_{\rm eV}$.  {\it Upper left:\/} Fluxes 
below the ionization potential of H at 13.6~eV, according to the model of \cite{Mathis1983}.  The energies that apply to the photoionization of neutral magnesium 
atoms are within the region bounded by the two vertical dashed lines.  The short 
purple line depicts the observed fluxes discussed in 
Section~\ref{sec:FUV_Rad_Field}.  {\it Lower colored traces:\/} Estimates for the 
unabsorbed local EUV and X-ray fluxes described in Section~\ref{sec:EUVX_fields} 
that are responsible for ionizing neutral H, He, N, O, and Ar, along with the ions of C, 
S, Mg, Si, and Fe.  The green lines at the bottom indicate the transmission factors for 
two values of the absorbing column densities of H~I.\label{fig:ISRF}}
\end{figure}

In order to derive $n(e)$ using Eq.~\ref{equilib1}, we must adopt a value for 
$I(E)_{\rm ISRF}$ that appears in Eq.~\ref{equilib2}.  There have been various 
estimates that have appeared in the literature for the strength of the ultraviolet 
photon flux in our location in the Galaxy, as summarized in Section 12.5 of \cite{Draine2011}.  We have chosen to use the fluxes vs. wavelength defined by \cite{Mathis1983}, which are 
depicted in Fig.~\ref{fig:ISRF} (black line in the upper left-hand corner). 

 We found 
that this radiation flux compares favorably with observations over a limited 
wavelength interval in the ultraviolet.  For instance, we evaluated the sum of two 
components that have been observed to contribute to $I(E)_{\rm ISRF}$.  One is the 
direct flux from stars at 1565\,\AA, where $I(\lambda)=1.21\, 10^{-6}\,{\rm 
erg~cm}^{-2}{\rm s}^{-1}{\rm \AA}^{-1}=9.53\, 10^4\,{\rm phot~cm}^{-
2}{\rm s}^{-1}{\rm \AA}^{-1}$ determined by \cite{Gondhalekar1980} from 
observations by the S2/68
Sky-survey telescope on the TD-1 satellite.  The other is the diffuse field of photons 
arising from the scattering of starlight by dust $I(\lambda)=3.8\, 10^4\,{\rm 
phot~cm}^{-2}{\rm s}^{-1}{\rm \AA}^{-1}$ over the interval 1370 to 1710\,\AA\ 
measured by \cite{Seon2011}\footnote{We avoided using the average fluxes from 
direct starlight measured by \cite{Seon2011} because their sky coverage omitted 
regions that contained particularly bright stars: e.g. compare Fig.~1 of \cite{Seon2011} with Fig.~1 of \cite{Gondhalekar1980} (but note that one depiction is 
inverted horizontally with respect to the other).}, who analyzed data from the  
SPEAR instrument on the Korean {\it STSAT-1}\ satellite mission.  The combination 
of these two fluxes is shown by the short purple line in Fig.~\ref{fig:ISRF}, which is almost imperceptibly above the flux level in the same wavelength region in the model of 
\cite{Mathis1983}.

\subsubsection{Atomic data for magnesium}\label{sec:atomic_data_Mg}

We adopted the photoionization cross sections $\sigma({\rm Mg}^0,E)$ calculated 
by \cite{Wang2010}, who used a fully relativistic $R$-matrix method.  These cross 
sections are in excellent agreement with the experimental results of \cite{Wehlitz2007}.  In our application of Eq.~\ref{equilib2} to these cross sections and our 
adopted ISRF ionizing flux (\S\ref{sec:FUV_Rad_Field}), we found that 
$\Gamma({\rm Mg}^0)=3.72\, 10^{-11}{\rm s}^{-1}$.\footnote{This 
ionization rate is lower than those adopted by \cite{Gry.Jenkins2001} and \cite{Slavin.Frisch2006}, which equaled $6.1~{\rm and}~4.5\, 10^{-11}{\rm s}^{-1}$, 
respectively.  The former used the radiation field of \cite{Mathis1983}, while the 
latter adopted the (stellar plus diffuse) field of \cite{Gondhalekar1980}.  Both 
authors used values of $\sigma({\rm Mg}^0,E)$ from the parametric description of 
\cite{Verner1996}, which is higher than the cross sections of \cite{Wang2010} at 
photon energies between the ${\rm IP(Mg^0)}$ and $10.5\,{\rm eV}$.}  Values for 
$C({\rm Mg}^+,T)$ have been calculated by \cite{Allan1988}, and they have been 
expressed in parametric form by \cite{Kingdon.Ferland1996}.  
For example, $C({\rm Mg}^+,6000\,{\rm K})=5.3\, 10^{-11}\,{\rm cm}^3{\rm s}^{-1}$.  
If the rate constant 
for the reaction ${\rm Mg}^0 + {\rm He}^+ \rightarrow {\rm Mg}^+ + {\rm He}^0$ 
is greater than a few times $10^{-10}\,{\rm cm}^3{\rm s}^{-1}$, this mode of 
ionizing Mg could make a non-negligible shift in the equilibrium.  Unfortunately, we 
were unable to find any determination (or even an estimate) for this rate in the 
literature.  For the contribution of radiative recombination to $\alpha({\rm 
Mg}^+,T)$, we adopted the analytic fit expressed by \cite{Badnell2006}.  For the 
additional effect from dielectronic recombination, we used the determinations 
published by \cite{Altun2006}.  At $T=6000\,$K, $\alpha({\rm 
Mg}^+,T)=1.4\, 10^{-12}\,{\rm cm}^3{\rm s}^{-1}$.

\subsubsection{Deriving constraints}\label{sec:constraints}

We equate $n({\rm Mg}^0)$ and $n({\rm Mg}^+)$ to our measurements of 
$N({\rm Mg~I})$ and $N({\rm Mg~II})$ and solve for $n(e)$ by re-expressing 
Eq.~\ref{equilib1} in the form,
\begin{equation}\label{equilib3}
n(e)={\Gamma({\rm Mg}^0)\over \alpha({\rm Mg}^+,T)[N({\rm Mg~II})/N({\rm 
Mg~I})]-C({\rm Mg}^+,T)/1.17}~,
\end{equation}
where the factor 1.17 is an approximate value for $n(e)/n({\rm H}^+)$ (the extra 
electrons come from the ionization of helium).

We recognize that the two column densities that appear in a quotient are subject to 
observational errors.   A conventional approach for deriving the error of a quotient 
is to add in quadrature the relative errors of the two terms, yielding the relative 
error of the quotient. However, this scheme breaks down when the error in the 
denominator is not very much less than the denominator itself.  For this reason, we 
resorted to a more robust way to derive the error of a quotient that has been 
developed by \cite{Geary1930} for two independent quantities whose errors are 
normally distributed; for a concise description of this method see Appendix~A of 
\cite{Jenkins2009}. We used this method here to define acceptable $\pm 1\,\sigma$ 
limits for the expression $N({\rm Mg~II})/N({\rm Mg~I})$ in Eq.~\ref{equilib3}.

\subsection{Fine-structure populations of ionized carbon atoms}\label{sec:CII_fsl}

The populations of the two fine-structure levels from the $J$-splitting of the ground 
state of ${\rm C}^+$, ${\rm level}~1=~^2{\rm P}_{1/2}$ and  ${\rm 
level}~2=~^2{\rm P}_{3/2}$, are governed by several excitation and de-excitation 
processes, the most important of which are collisions with electrons with rate 
constants $\gamma_{1,2}(e,T)$ (upward) and $\gamma_{2,1}(e,T)$ (downward), 
which are balanced against spontaneous decays of the upper level radiating at 
$\lambda=157.6\,\mu$m with an Einstein $A$-coefficient $A_{2,1}=2.29\,
10^{-6}\,{\rm s}^{-1}$ \cite{Nussbaumer.Storey1981}.  Additional channels which 
are small in comparison but not completely negligible include collisions with 
protons and neutral hydrogen atoms [rate constants $\gamma_{1,2}({\rm H}^+,T)$ 
and $\gamma_{1,2}({\rm H}^0,T)$ and their de-excitation counterparts] and optical 
pumping [rates $r_{1,2}({\rm OP})$ and $r_{2,1}({\rm OP})$] by the ISRF acting on 
the two strong transitions out of the ground state at 1036 and 1335\,\AA.  
Balancing all rates, we find that
\begin{eqnarray}\label{fsl_equilib}
&&[n(e)\gamma_{1,2}(e,T) + n({\rm H}^+)\gamma_{1,2}({\rm H}^+,T)\nonumber\\
&+& n({\rm H}^0)\gamma_{1,2}({\rm H}^0,T) + r_{1,2}({\rm OP)}]n({\rm 
C}_{1/2}^+)\nonumber\\
&=&[n(e)\gamma_{2,1}(e,T) + n({\rm H}^+)\gamma_{2,1}({\rm H}^+,T)\nonumber\\
&+& n({\rm H}^0)\gamma_{2,1}({\rm H}^0,T) +  r_{2,1}({\rm OP})+A_{2,1}]n({\rm 
C}_{3/2}^+)~.
\end{eqnarray}
For the electron, proton and hydrogen rate constants, we know from the principle of 
detailed balance that
\begin{equation}\label{detailed_balance}
\gamma_{1,2}=(g_2/g_1)\exp\left( -{E\over kT}\right) \gamma_{2,1}~,
\end{equation}
where $g_1=2$, $g_2=4$, and $E/k=91.2\,$K.  The downward rate constant for 
electron collisions is given by
\begin{equation}\label{g21(e,T)}
\gamma_{2,1}(e,T)={8.63\, 10^{-6}\Omega(e,T)\over g_2T^{1/2}}~,
\end{equation}
where we have adopted the collision strengths as a function of temperature 
$\Omega(e,T)$ from a fit by \cite{Draine2011} to the results of \cite{Tayal2008}.  Rates for 
excitations by protons $\gamma_{1,2}({\rm H}^+,T)$ have been calculated by \cite{Bahcall.Wolf1968} for $T>10^4\,$K; below this temperature such excitations are 
negligible compared to those caused by electrons.  For $\gamma_{2,1}({\rm H}^0,T)$ 
we used the analytic fit given by \cite{Barinovs2005}, and once again we can use 
the principle of detailed balance to determine $\gamma_{1,2}({\rm H}^0,T)$.  We 
calculated that $r_{1,2}({\rm OP})=1.35\, 10^{-10}{\rm s}^{-1}$ and 
$r_{2,1}({\rm OP})$ is half as large.  As stated earlier, we assumed that $n({\rm 
H}^+)\approx n(e)/1.17$ and $n({\rm H}^0)\approx 0.2\,{\rm cm}^{-3}$;  errors 
in these assumptions will not be important because the effects of proton and 
hydrogen collisions are very small compared to those from electrons.

To solve for $n(e)$ we rewrite Eq.~\ref{fsl_equilib} in the form
\begin{eqnarray}\label{n(e)_solve}
n(e)&=&\Big\{ [n({\rm H}^0)\gamma_{2,1}({\rm H}^0,T) + r_{2,1}({\rm OP})+ 
A_{2,1}][n({\rm C}_{3/2}^+)/n({\rm C}_{1/2}^+)]\nonumber\\
&-& n({\rm H}^0)\gamma_{1,2}({\rm H}^0,T) - r_{1,2}({\rm 
OP})\Big\}\Big/\nonumber\\
&& \Big\{\gamma_{1,2}(e,T) + \gamma_{1,2}({\rm H}^+,T)/1.17\nonumber\\
&-& [\gamma_{2,1}(e,T) + \gamma_{2,1}({\rm H}^+,T)/1.17][n({\rm 
C}_{3/2}^+)/n({\rm C}_{1/2}^+)]\Big\}
\end{eqnarray}
Again, we used Geary's (1930) method to evaluate the error in a quotient, in this 
case for $n({\rm C}_{3/2}^+)/n({\rm C}_{1/2}^+)\equiv N({\rm C~II}^*)/N({\rm 
C~II})$, as we had done earlier for $N({\rm Mg~II})/N({\rm Mg~I})$.  We did not 
employ this scheme for sensing the uncertainty of the numerator over the 
denominator in Eq.~\ref{n(e)_solve} because the errors of the two are correlated 
instead of being independently random.

\subsection{Combining the Mg and C constraints:  results for n(e) and T}\label{sec:MgC}
\begin{figure}
\includegraphics[width=1.0\columnwidth]{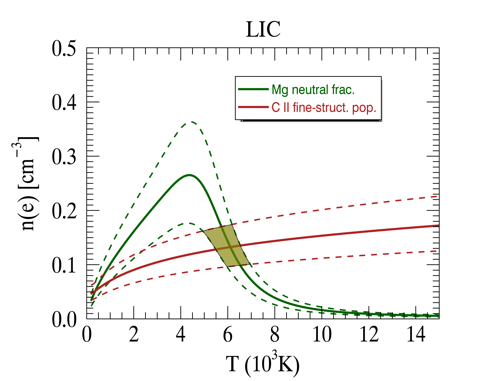}
\caption{Allowed values for $n(e)$ and $T$ defined through the use of 
Eq.~\ref{equilib3} for the observation of $N({\rm Mg~I})/N({\rm Mg~II})$ (solid 
green line) and Eq.~\ref{n(e)_solve} for $N({\rm C~II}^*)/N({\rm C~II})$ (solid 
red line) for the component identified as the LIC toward $\alpha$~Leo.  Dashed line 
counterparts indicate the $\pm 1\,\sigma$ envelopes arising from errors in the 
column densities.  The olive-colored fill indicates the region that satisfies both types 
of measurement within their $1\,\sigma$ uncertainties.\label{fig:MgC_LIC}}
\end{figure}

We can derive a unique combination of $n(e)$ and $T$ (and error limits thereof) by 
plotting the outcomes of Eqs.~\ref{equilib3} and \ref{n(e)_solve} on a diagram for 
these two quantities and finding the intersection of the two curves.  The outcome is 
shown in 
Fig.~\ref{fig:MgC_LIC} for the component at V~=~8.8\,\kms\ identified as the LIC. For this component we find T = 6000\,K $\pm$1000\,K in very good agreement with the outcomes from the line fits, and 
$n(e)=0.13(+0.04,-0.035)\,{\rm cm}^{-3}$.

\begin{figure}
\includegraphics[width=1.0\columnwidth]{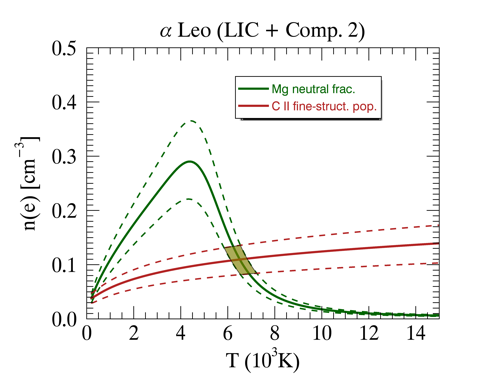}
\caption{As  Fig.~\ref{fig:MgC_LIC} for the sum of both velocity components toward $\alpha$~Leo. \label{fig:MgC}}
\end{figure}
Fig.~\ref{fig:MgC} shows the outcome for the sums of column densities of both velocity 
components.  The use of the sum of column densities is justified since the column density distribution among both components does not vary much from species to species 
as shown in the last column of Table~\ref{tab:results}, which indicates that both components are likely to experience similar conditions. The closeness of the outcomes of Figs.~\ref{fig:MgC_LIC} and \ref{fig:MgC}  further validates this assumption.
 We find that for the sum of both components $n(e)=0.11(+0.025,-0.03)\,{\rm cm}^{-3}$. 
The preferred temperature of 6500$\pm$700\,K derived with the sum of column densities is slightly higher than the 
outcomes from the line fits, but $T=6000\,$K is still allowed within the $1\sigma$ 
error deviations. \footnote{Both Fig.~\ref{fig:MgC_LIC} and Fig.~\ref{fig:MgC}, allow another low-temperature solution, which we however do not consider since it would be in contradiction with the temperature results from the fits.}

\section{Ionization balance of other elements}\label{sec:ionization_other} 
Ultimately, we will want to make an estimate for the density of neutral and ionized 
hydrogen by developing a model for the partial ionization of various atomic species 
and comparing the results with our observations.  To solve for equilibrium 
ionization fractions, we must balance the ionization rates against the effects of 
radiative and dielectronic recombination with free electrons.  Photons with energies 
above the ionization potential of hydrogen (13.6\,eV) provide the principal means 
for governing the ionization balance between the preferred and next higher ionization stages of atoms under 
consideration.  Charge exchange reactions also play a role in modifying the 
ionization balances.  A representative ionization rate of a few$\, 10^{-16}{\rm 
s}^{-1}$ by cosmic rays \citep{Indriolo.McCall2012, Indriolo2015} is more than 
two orders of magnitude below our computed ionization rates by photons and thus 
can be ignored.  The basic equations that one can use to evaluate the equilibrium 
concentrations of different ionization stages in a partially ionized medium have 
been outlined by \cite{Jenkins2013}\footnote{The relevant equations in \cite{Jenkins2013}
are those given in Sections~6.4 and 6.5 of that paper, which apply to elements that 
have appreciable amounts of their atoms in neutral form.  For the equilibrium of any 
predominantly singly ionized element X, the fractions $f_0({\rm X}, T)$,  $f_+({\rm 
X}, T)$ and $f_{++}({\rm X}, T)$ can be evaluated using the same equations and 
rates for the higher stages of ionization, except that these three fractions then apply 
to the concentrations of X$^+$, X$^{++}$, and X$^{3+}$, respectively.}.  In the 
subsections that follow, we describe the properties of the principal parameters that 
we adopted for computing the equilibria.

\subsection{Extreme Ultraviolet and X-ray radiation fields}\label{sec:EUVX_fields}

For energies above 13.6\,eV, nearly all of the Galactic direct and scattered radiation 
from stellar sources considered in Section~\ref{sec:FUV_Rad_Field} is blocked by 
the opacity of distant clouds of neutral hydrogen.  For the special circumstances 
where this blockage is small, such as for nearby white dwarf stars or the B-type 
stars $\epsilon$ and $\beta$~CMa, the local region can be exposed to energetic 
radiation \citep{Craig1997}, although it is much weaker than the fluxes at lower 
energies.  The blue line in Fig.~\ref{fig:ISRF} shows a summation of fluxes from 
these sources at the Earth's location computed by \cite{Vallerga1998}, after we have
de-absorbed the radiation by multiplying it by $\exp[\tau({\rm H}^0,E)]$ for a 
representative column density $N({\rm H}^0)=9\, 10^{17}\,{\rm cm}^{-2}$ to 
the edge of the LIC \citep{Wood2005, Gry.Jenkins2014}.  Stars are the 
dominant source of ionizing radiation at energies just below the ionization potential 
of hydrogen, but their output falls below other radiation sources at higher energies.

Soft X-rays that could influence the local ionization are emitted by hot gases in three 
principal domains: (1) distant regions in the disk and halo of our Galaxy, 
characterized by the non-local emission defined by \cite{Kuntz.Snowden2000}, 
which they labeled as the trans-absorption emission (TAE), (2) emission from 
$10^6\,$K gas in the Local Bubble, and (3) emission from an interface between the 
warm cloud and the surrounding hot medium (but see our conclusions in Section~\ref{sec:interface}; perhaps this interface radiation is weaker).  \cite{Jenkins2013} has derived the fluxes 
at different energies for the distant X-ray background based on the description of 
TAE provided by \cite{Kuntz.Snowden2000}.  These fluxes are shown by the orange 
trace in Fig.~\ref{fig:ISRF}.  Radiation levels from the hot gas in the Local Bubble 
and the LIC boundary have been estimated by \cite{Slavin.Frisch2008} 
(respectively shown as brown and red traces in the figure).  However, \cite{Slavin.Frisch2008} made no correction to the apparent Local Bubble emission to account 
for the fact that the soft X-ray background measurements were contaminated by the 
emission of X-rays caused by charge exchange when the solar wind interacts with 
exospheric hydrogen in the Earth's magnetosheath and the heliospheric 
contribution arising from interactions with incoming interstellar neutral H and He.   
To correct for these two sources of emission, we have multiplied the apparent hot 
gas emission by a factor of 0.6 \citep{Galeazzi2014,Snowden2014, Snowden2015}, 
which now serves as our estimate for the fraction of the observed diffuse flux that is 
produced by hot gas in the Local Bubble. 

There are significant uncertainties in calculating the strength of the ionizing 
radiation for atoms and ions situated in the sight line toward $\alpha$~Leo.  First, 
the unabsorbed flux intensities discussed earlier are approximate.  Second, we 
assume that virtually all of the gas experiences the same amount of exposure to this 
radiation, i.e., we do not attempt to consider that some gas near the edge of the 
cloud is irradiated more strongly than that at the center.  Finally, we describe the 
attenuation of this flux interior to our gas region in terms of a simple
one-dimensional absorption law without considering any details of the complex 
(and unknown) geometry of the cloud or any full treatment of the three-dimensional 
radiative transfer.  For the most part, errors that arise from neglecting these 
complications will influence the outcome for what we designate as a representative 
absorbing column density, which itself is a free parameter that we adjust to give a 
satisfactory agreement between our calculations and the observed column densities.  
Also, our calculations of the influence of absorption in changing the ionization 
balance is important only for the elements H, N, O, and Ar, and has only a weak 
influence on C, S and Mg.  The species Si and Fe remain almost entirely in the 
preferred singly ionized state, regardless of any reasonable changes in the radiation 
field strength.

For interstellar clouds with significantly higher self-shielding of the external 
radiation, the production of photons by internal sources, such as the chromospheric 
emissions by embedded late-type stars, can be relatively important.  For our case 
with the sight line to $\alpha$~Leo, the shielding is so low that radiation from 
external sources completely overwhelms that from the internal 
ones.\footnote{While some white dwarfs are inside the LIC, their radiations 
are included in the sum of white dwarf fluxes compiled by \cite{Vallerga1998}.}

\subsection{Secondary ionization processes}\label{sec:secondary_ionization}

In the appendices of \cite{Jenkins2013}, there are descriptions of a number of 
secondary ionization processes that can in principle provide additional ionization 
routes that supplement the primary ones.  However, owing to the fact that most of 
the ionizations are caused by photons with energies that are not far above 13.6\,eV, 
the relative influence of these secondary processes is quite small.  For instance, 
there are very few electrons produced by the photoionization of H$^0$, He$^0$, 
and He$^+$ that have sufficient energy to create secondary ionizations of H$^0$ 
and He$^0$.  The only non-primary modes that are not negligible in our case are the 
emissions of photons arising from the recombinations of He$^+$ and He$^{++}$ 
with electrons to produce He$^0$ and He$^+$.  For instance, the rates 
$\Gamma_{\rm He^0}({\rm H}^0)$ and $\Gamma_{\rm He^+}({\rm H}^0)$ 
respectively represent 14\% and 4\% of the total H ionizations.

\subsection{Atomic physics properties}\label{sec:atomic_physics}
 \subsubsection{Ionization cross sections}\label{sec:cross_sections}

For H$^0$ and He$^+$, we use the cross sections described by \cite{Spitzer1978} (pp. 
105-106).  Cross sections for He$^0$ and Ar$^0$ are taken from \cite{Samson1994} and \cite{Marr.West1976}, respectively.  Cross sections vs. energy for other 
neutral, singly and doubly ionized species are taken from the fitting formulae 
provided by \cite{Verner.Yakovlev1995}, which describe not only the cross sections 
for ionizing the outer electron, but also those for inner shell electrons by photons at 
higher energies.  

\subsubsection{Radiative and dielectronic recombination coefficients}\label{sec: 
recombination_coeff}

Radiative recombinations to form H$^0$ and He$^+$ are taken from \cite{Spitzer1978} (pp. 
105-107), and we used the radiative and dielectronic rates of \cite{Aldrovandi.Pequignot1973} for the creation of He$^0$.  Recombinations to the lowest 
electronic level of H generate Lyman limit photons, most of which are absorbed 
within the region to ionize some other H atom.  Thus, we excluded recombinations 
to this level by employing the Case~B recombination coefficient \cite{Baker.Menzel1938}.  Fitting formulae for the recombinations of other elements were taken from 
\cite{Shull.VanSteenberg1982}, supplemented by rates to low lying levels of N$^0$, 
N$^+$, C$^+$, C$^{++}$ \citep{Nussbaumer.Storey1983} and Si$^+$ \citep{Nussbaumer.Storey1986}.

\subsubsection{Charge exchange reaction rates}\label{sec:charge_exchange}

For computing the charge exchange rates with H$^0$ for all elements except O, we 
used the fits specified by \cite{Kingdon.Ferland1996}.  For those elements whose 
ionizations can be reduced by capturing an electron from He$^0$ (but again 
excluding O), we evaluated rate constants for $T=6000\,$K by making logarithmic 
interpolations between the results for $T=10^{3.5}$ and $10^4\,$K given by 
\cite{Butler.Dalgarno1980}.  Charge exchange rate constants for O$^+$ reacting with 
H$^0$ and He$^0$ are respectively from \cite{Stancil1999} and \cite{Butler1980}.

\subsection{Column density of H~I and H~II}\label{sec:N(Htot)}

Our observation of the total column density of nitrogen, $N({\rm N_{tot}})=N({\rm 
N~I})+N({\rm N~II})$ is our most secure way to estimate the total amount of 
neutral and ionized hydrogen in the sight line (our computed abundance of doubly 
ionized N should be very low, see Table~\ref{tbl:ion_fractions} in 
Section~\ref{sec:col_dens_other}).  Generally, the depletion of nitrogen in the ISM is 
quite weak ($-0.109\pm 0.119$~dex), and it does not seem to vary when the 
depletions of other elements change \citep{Jenkins2009}\footnote{Any uncertainty in 
the reference (solar) abundance adopted by \cite{Jenkins2009} should not affect our 
inferred factor to multiply $N({\rm N_{tot}})$ to get $N({\rm H_{tot}})$ since 
changes of opposite sign would be reflected in the derived amount of depletion.}.  
We find that 
 for the full line of sight, e.g. the sum of both velocity components,
$N({\rm H_{tot}})=2.83\,(+1.18,-0.69)\,10^{18}{\rm cm}^{-2}$.  
Our specified error range for this column density includes the uncertainty in the sum of 
two observed nitrogen column densities and the uncertainty in the depletion, both 
of which are combined in quadrature. 
 In this case we only consider the sum of the components since Copernicus data do not allow us to derive precise \NII\ column densities for each component separately.

For the partition of $N({\rm H_{tot}})$ into the expected values of $N({\rm H^0})$ 
and $N({\rm H^+})$, we must rely 
chiefly on the observed nitrogen ionization fraction $N({\rm N~I})/N({\rm N_{tot}})=0.60(+0.042,-0.123)$ combined with 
our model for the photoionization of 
hydrogen.  The most stringent way to confine the free parameters in this model is to 
force the outcome for $n(e)$ to fall within the range of the observational result that 
we specified in Section~\ref{sec:MgC}. 
 In making use of $n(e)$, we acknowledge that it is driven by two quantities that are not measured explicitly: (1) the local volume density of hydrogen and (2) the amount of shielding of the external radiation by hydrogen and helium (as illustrated by the green traces in Fig.~\ref{fig:ISRF}).  By adjusting these two parameters, we can explore combinations that are consistent with our observed quantities, the nitrogen ionization fraction and $n(e)$.  
 Using 
this technique, we find that a representative local density $n({\rm 
H_{tot}})=0.30(+0.10,-0.13)\,{\rm cm}^{-3}$ gives a satisfactory fit, where the 
shielding of the external radiation could be within the range $4.3-7.0\,
10^{17}{\rm cm}^{-2}$ of neutral hydrogen (and with $n({\rm He}^0)/n({\rm 
H}^0)=0.07$).\footnote{The largest value expressed here does not appear in 
Table~\ref{tbl:outcomes} because it pertains to $[n(e),n({\rm 
H_{tot}})]=[0.08,0.25]$, which is an acceptable combination.} The model indicates 
that the neutral fractions of H range from 0.58 to 0.73.

The upper portion of Table~\ref{tbl:outcomes} presents the outcomes for the expected column densities 
of H$^0$ (along with other elements to be discussed later).  The three columns that 
show such column densities present values that pertain to three combinations of 
$n({\rm H_{tot}})$, $n(e)$, and shielding column densities that gave acceptable 
results.  For each column, the three rows labeled ``Upper,'' ``Best,'' and 
``Lower'' indicate the values of $N({\rm H^0})$ that are allowed within the 
uncertainties of $N({\rm H_{tot}})$. 

\subsection{\HI\ column density -- A note of caution on the use of Lyman $\alpha$ to derive  $N$(\HI)  }\label{sec:HI}
\begin{figure}
\includegraphics[width=1.0\columnwidth]{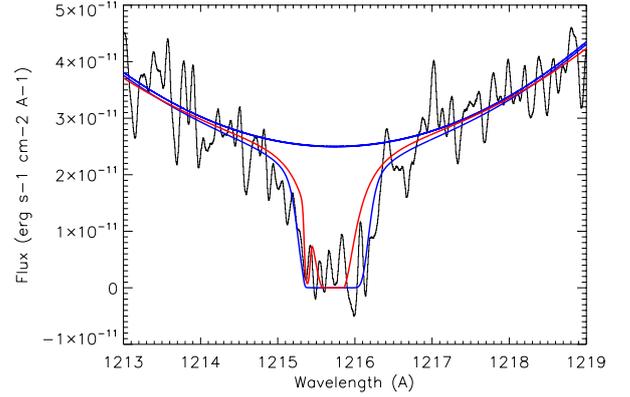}  
\caption{Lyman~$\alpha$ profile in the spectrum of \aleo. The original data have been smoothed by 20 pixels to reduce the apparent noise.  The red trace shows the expected profile from the two warm components with a total $N({\rm H~I})_{tot}=1.9~10^{18}\,{\rm cm}^{-2}$ (as favoured by our model).   The blue trace shows the profile when an extra component   with T= 200\,000 K and 
 $N$(\HI) = 2.1\,$10^{15}  {\rm cm}^{-2}$ 
is added to the fit. (see section~\ref{sec:interface}). In all models an absorption of \DI\ is included,with a fixed D/H ratio of 1.6\,10$^{-5}$. \label{fig:HI}}
\end{figure}

There is a large discrepancy between the apparent   $N$(\HI) derived from the Lyman $\alpha$ profile assuming the 2-component line-of-sight structure ($N$(\HI) = 6.6\,10$^{18}$\,cm$^{-2}$) and the total   $N$(\HI) derived from the ionization model calculations (between 1.3 and 2.8 \,10$^{18}$\,cm$^{-2}$, as shown in Table~\ref{tbl:outcomes}). 
Despite the high level of noise in the spectrum, Fig.~\ref{fig:HI} shows that the low column density from the model (red trace) does not produce a good fit to the data. On the other hand, as high a column density as 6.6\,10$^{18}$\,cm$^{-2}$ would imply the existence of extraordinarily large depletions for many metals, ones that are even higher than those that we discuss later (Section~\ref{sec:col_dens_other}) and that approach the highest strengths of depletion  in the Milky Way. 
This is a prospect that is very unlikely in such a diffuse gas. 

The danger of overestimating   $N$(\HI) from the Lyman $\alpha$ line 
 when $N$(\HI) is lower than about 10$^{19}$\,cm$^{-2}$
had already been discussed by \cite{VidalMadjar.Ferlet2002} and is supported further by the recognition by  \cite{Wood2005} about the presence of extra absorption by hot gas from the heliosphere and/or an astrosphere around a number of target stars. Here the discrepancy may be due to the presence of very small amounts of high-velocity or high-temperature gas  present in the line of sight and undetectable in the lines of other elements. 
This possibility will be brought up again when we discuss the presence of interfaces with the hot gas in Section~\ref{sec:interface}.

Another interesting possibility is related to the claim by \cite{Gies2008} that \aleo\ has a previously undiscovered close companion, which may be a white dwarf star. \cite{Gies2008}  suggest that although this companion should be much fainter than \aleo, it may contribute a non-negligible flux in the FUV and they note that "in fact, \cite{Morales2001} find that the spectral energy distribution of Regulus is about a factor of 2 brighter in the 1000 $-$ 1200 \AA\ range than predicted by model atmospheres for a single B7 V star". In the context of this possibility, we conjecture that the Lyman-$\alpha$ absorption from the white dwarf atmosphere could be responsible for the extra absorption in the wings of the interstellar profile.  However this hypothesis cannot be investigated any further because of the insufficient signal in the Lyman-$\alpha$ core, as well as the lack of information on the potential close companion, which has not yet  been detected directly \cite{Absil2011}.
\subsection{Predictions for the column densities of other 
elements}\label{sec:col_dens_other}
 \begin{table}
\caption{Predicted Column Densities for The Sum of Both Components$^a$ \label{tbl:outcomes}}
\begin{tabular}{
l	
c	
c	
c	
c 
}
               \hline 
Atom&Lower & Preferred& Upper&\\
or Ion & Extreme & Value & Extreme&\\
               \hline 
                \multicolumn{5}{c}{Local Densities and Shielding of Radiation}\\
               \hline 
$n({\rm H_{tot}})({\rm cm}^{-3})$&0.17&0.30&0.40&\\[2pt]
$n({\rm H^0})({\rm cm}^{-3})$&0.10&0.20&0.28&\\
$n({\rm H^+})({\rm cm}^{-3})$&0.07&0.10&0.12&\\[2pt]
  $n({\rm He^0})({\rm cm}^{-3})$&$7.2(-3)$&0.016&0.023&\\
$n({\rm He^+})({\rm cm}^{-3})$&$9.3(-3)$&0.014&0.017&\\
$n({\rm He^{++}})({\rm cm}^{-3})$&$5.2(-4)$&$5.3(-4)$&$5.2(-4)$&\\[2pt]
$n(e) ({\rm cm}^{-3})$&0.08&0.11&0.135&\\[2pt]
Shielding &&&&\\
$N({\rm H^0})({\rm cm}^{-2})$&4.25(17)&4.00(17)&3.50(17)&\\

               \hline 
 \multicolumn{4}{c}{Expected Column Densities$^b$}&Observed\\
               \hline 
Upper $N$(H$^0$)&2.4(18)&2.7(18)&2.8(18)&\\
Best $N$(H$^0$)&1.7(18)&1.9(18)&2.0(18)&\\
Lower $N$(H$^0$)&1.3(18)&1.4(18)&1.5(18)&\\[3pt]
Upper $N$(C$^+$)&7.7(14)&7.8(14)&7.9(14)&8.1(14)\\
Best  $N$(C$^+$)&5.2(14)&5.3(14)&5.3(14)&6.7(14)\\
Lower $N$(C$^+$)&3.7(14)&3.8(14)&3.8(14)&5.9(14)\\[3pt]
Upper $N$(O$^0$)&1.0(15)&1.2(15)&1.2(15)&1.3(15)\\
Best $N$(O$^0$)&7.2(14)&8.2(14)&8.4(14)&1.15(15)\\
Lower $N$(O$^0$)&5.3(14)&6.0(14)&6.2(14)&1.02(15)\\[3pt]
Upper $N$(Mg$^+$)&1.4(13)&1.6(13)&1.7(13)&6.7(12)\\
Best $N$(Mg$^+$)&9.8(12)&1.1(13)&1.2(13)&6.5(12)\\
Lower $N$(Mg$^+$)&7.4(12)&8.4(12)&8.9(12)&6.3(12)\\[3pt]
Upper $N$(Si$^+$)&2.0(13)&2.0(13)&2.0(13)&1.36(13)\\
Best $N$(Si$^+$)&1.4(13)&1.4(13)&1.4(13)&1.30(13)\\
Lower $N$(Si$^+$)&1.1(13)&1.1(13)&1.1(13)&1.24(13)\\[3pt]
Upper $N$(S$^+$)&8.0(13)&8.3(13)&8.4(13)&2.68(13)\\
Best $N$(S$^+$)&4.5(13)&4.6(13)&4.7(13)&2.44(13)\\
Lower $N$(S$^+$)&2.6(13)&2.7(13)&2.8(13)&2.20(13)\\[3pt]
Upper $N$(Ar$^0$)&2.0(12)&2.7(12)&3.0(12)&1.93(12)\\
Best $N$(Ar$^0$)&1.3(12)&1.8(12)&2.0(12)&1.86(12)\\
Lower $N$(Ar$^0$)&9.2(11)&1.2(12)&1.4(12)&1.79(12)\\[3pt]
Upper $N$(Fe$^+$)&2.6(12)&2.6(12)&2.6(12)&2.48(12)\\
Best $N$(Fe$^+$)&1.9(12)&1.9(12)&1.9(12)&2.45(12)\\
Lower $N$(Fe$^+$)&1.4(12)&1.4(12)&1.4(12)&2.42(12)\\
               \hline 
               \end{tabular}\\
{\it $^a$} Notation: A(B) refers to a column density ${\rm A}\times 
10^{\rm B}\,{\rm cm}^{-2}$.\\
{\it $^b$} Includes the effects of ionization and depletion (for $F_*=0.6$).

\end{table}

Using the radiation fields and atomic data outlined in 
Sections~\ref{sec:EUVX_fields} to \ref{sec:atomic_physics}, we can compute the 
expected column densities for various atomic species that we were able to observe 
(other than N, which was used to predict the amount of H and its ionization 
fractions).  
In the first attempt to do this we 
assumed that their abundances equal the solar values relative to 
hydrogen and that their distribution of ion stages followed our ionization model.  
This revealed that lightly depleted species, such as C, O, 
 and Ar, 
gave a good agreement with the observations, but elements that are known to be 
depleted in the ISM showed predicted values considerably above the observed ones.  

We now will investigate whether or not the pattern of depletions follows trends for 
different elements seen elsewhere in the local part of our Galaxy.  In addition, we 
have a goal of understanding where the gas toward $\alpha$~Leo stands within the 
scale of severity of depletions.  To accomplish these tasks, we make use of the 
parametric description developed by \cite{Jenkins2009} that defines the depletions in 
terms of some constants that apply to each element and a variable known as $F_*$ for the 
overall strength of depletions from one location to the next.  The elements S and Ar 
have undetermined or probably small depletions: we 
 first
assume zero depletions in our 
comparisons of computed column densities against observed ones,
 however these comparisons will evidence that  a non-zero depletion is required for S (Section~\ref{sec:depletions}).

We have adjusted $F_*$ to give the least discrepancy between
the calculated and observed abundances of the most depleted elements Mg, Si and Fe. 
After considering the combined errors 
(added in quadrature) 
in the observations, our determination of 
$N({\rm H_{tot}})$, and the depletion constants for each element, 
we find that we have a good match
 with the observations for all elements except Mg when we 
specify a value of $F_*$ equal to 0.6.  
This value for $F_*$ seems extraordinarily 
high for a low density medium.  Mg appears to be more strongly depleted than what 
we would expect for our most favorable $F_*$, as the upper error limit for the 
observed column density of Mg~II does not overlap any of the lower limits for the 
predictions for $N({\rm Mg^+})$.  
Table~\ref{tbl:outcomes} shows the outcomes 
for our models, which may be compared with the observations of the total column 
densities of both velocity components 
from

 the fourth column of Table~\ref{tab:results}.
  \section{Interpretation: Implications for the properties of the Local ISM}\label{sec:interpretation}
We have shown in the previous section that 
  when combined with a comprehensive ionization model, our extensive collection of column densities for many different species
in the direction of \aleo\  has offered a unique opportunity for us to develop a well constrained and internally consistent description for the state of the gas.
This development allows us to surpass most previous analyses of the nearby interstellar material around the Sun.
 \subsection{Temperature}
We have estimated  temperatures by two independent methods. The first estimate of $T$  directly results from the line profile analysis, as explained in Section~\ref{sub:line-fitting}, and as listed in Table~\ref{tab:results}. The second method is the analysis shown in Figs~\ref{fig:MgC_LIC} and \ref{fig:MgC}, which combines constraints given by the  ionization equilibrium of magnesium and the fine-structure populations of ionized carbon.
 We have applied the latter method considering first the two velocity components as a whole, and we find  a gas temperature of 6500$\pm$700\,K. We also considered the column densities derived for  the LIC component alone, and we find a somewhat broader temperature range, $T=6000\pm$1000\,K.  This temperature is in remarkable agreement with the temperature derived from the line profile analysis, $T=6000\pm$600\,K.

 Our temperature estimate is consistent with the weighted mean gas temperature derived by \cite{Redfield.Linsky2004b} in the local warm clouds, found equal to 6680 K with a dispersion of 1490 K.

Note that it is similar although only marginally consistent with the temperature derived from IBEX outside the heliosphere ($T\sim\,$7500 K ; \cite{McComas2015}. The difference may illustrate varying physical conditions within the LIC.
\subsection{Densities}
The electron density $n(e)=0.13(+0.04,-0.035)\,{\rm cm}^{-3}$ that we derive 
 from Fig.~\ref{fig:MgC_LIC} 
for the LIC alone is in excellent agreement with the range 0.125 $\pm 0.045$ derived for the LIC in the direction to $\epsilon$ CMa by \cite{Gry.Jenkins2001} as well as with the weighted mean value of $n(e)=0.12\pm0.04\,{\rm cm}^{-3}$ derived  in 7 lines of sight by \cite{Redfield.Falcon2008}. 

 The total hydrogen density $n(H_{tot})= n(H^0) +n(H^+)$ derived  from n(e) and the ionization fraction of N via our ionization model, is  $n(H_{tot})=0.3(+0.1,-0.13)\,{\rm cm}^{-3}$.

The neutral hydrogen density then defined by the neutral fractions allowed within our model is

$n(H^0)=0.20(+0.08,-0.10)\,{\rm cm}^{-3}$. 
This value is in agreement with the value customarily cited in cloud modeling e.g. as in \cite{Slavin.Frisch2008}. 
It is also supported by the measurement of $N$(\HI) toward AD Leo, 8.5\degree\
away in the sky from \aleo\ and only 4.7 pc away from the Sun. For AD Leo,  \cite{Wood2005} reported  log $N$(\HI) = 18.47, which seems surprisingly high
for such a nearby star.  The implied average density over that sight line is $n$(\HI)= 0.20 \cmc\, which is
the same as the preferred $n$(\HI) derived from the ionization calculation for \aleo.

However, we note that this density contrasts with the mean value $n$(\HI) = 0.053 \cmc\ derived by GJ14 for the average of $N$(\HI)/$d$ in the four lines of sight in the $N$(\HI) sample of \cite{Wood2005}, where the presence of astrospheres around the target stars suggested that the sight lines are completely filled. We had noted that $n$(\HI) varies in the LIC since the outcomes spanned values from 0.03 to 0.1 \cmc. Our new value for $n$(\HI)  in the line of sight toward \aleo, derived independently by a different method, is  higher than all previous values and could be the evidence of an even larger $n$(\HI) variation inside the LIC.

The total gas density $n= n(H_{tot}) + n(e) + n(He) = 0.44(+0.13, -0.17)\,{\rm cm}^{-3}$.
 
\subsection{Pressure}  
The thermal pressure of the gas, 
calculated from T = 6500$\pm$700\,K and the above total gas density n, $\log(p/k)=3.46(+0.12,-0.22)$, 
is consistent with 
pressures found elsewhere in our part of the Galaxy, $\log(p/k)=3.58$ with an rms 
dispersion of 0.18\,dex \citep{Jenkins.Tripp2011}.  This pressure is substantially 
higher than the turbulent pressure $\rho b_{\rm turb}^2/2 = 1.1\,
10^{-14}\,{\rm dyne~cm}^{-2}$ ($\log (p/k)=1.91$), but it is considerably
lower than that derived by  \cite{Snowden2014} for the surrounding X-ray emitting 
hot gas, $\log(p/k)=4.025(-0.046,+0.42)$.  
This pressure difference between the two media might be 
reconciled by the support provided by a magnetic field within the LIC. 

\cite{Zirnstein2016} derived from the observation of the IBEX ribbon a magnitude of 
B=$2.93 \pm0.08\,\mu {\rm G}$ for the local interstellar magnetic field  far (1000 au) from the Sun. This magnetic field strength is equivalent to a pressure of only 2500 \cmc K, which, if not stronger elsewhere, is insufficient to allow the LIC to withstand the pressure value for the surrounding hot gas. With Voyager~I data however, \cite{Burlaga.Ness2014} measured in 2013 a varying interstellar magnetic field of average $4.86\,\mu {\rm G}$ with a dispersion of 0.45\,$\mu {\rm G}$, which approaches the value $\sim 5\,\mu {\rm G}$  needed to sustain a pressure balance with the hot gas. We will revisit this issue in Section~\ref{sec:interface})
\subsection{Filling factor} \label{sec:filling}
On the assumption that the gas has a uniform density in  the region responsible for the two absorption components and a 
considerably lower density elsewhere, we find that the volume of the partially neutral gas that 
intercepts the sight line toward $\alpha$~Leo is small, i.e., we obtain a filling factor 
equal to $N({\rm H^0})/[n({\rm H^0})d]=0.13(+0.19,-0.04)$, where $d=24\,$pc. It follows that in this direction the extent of the partially neutral gas that includes the LIC  is somewhere in the interval between 2.2 and 7.7 pc. 
 
The remaining 16 to 22 pc is thus devoid of detectable amounts of neutral or partially neutral gas. The question   is what fills the remaining space, which amounts to asking what fills the Local Bubble. \cite{Snowden2015} attribute 
70 $\pm22$ RU (Rosat Units) of the ROSAT 1/4\,kev emission detected in the direction to the LLCC
to the emission of the hot bubble gas in the foreground to the
cold cloud (after a correction of the ROSAT 1/4\,kev data for the heliospheric and magnetosheath Solar Wind Charge eXchange (SWCX) emission). They interpret this emission as occurring over a path length of 29 $\pm$ 11 pc in the Local Bubble plasma, made of gas at $T= 1.18\pm0.01~10^6$\,K and $n(e)=4.68\pm0.47\,10^{-3}$\cmc\  \citep{Snowden2014}. If this hot, soft X-ray emitting, gas fills most of the line of sight to the LLCC, it is more than probable that it also fills a large fraction of the line of sight to \aleo, which is  only 4\degree\ away.

We can note therefore that the star \aleo\ at a distance of 24 pc is very likely to be surrounded by hot gas. We have no evidence that it is embedded in partially neutral gas and therefore we do not expect it to exhibit an astrosphere. Thus the extra absorption in the Lyman-$\alpha$  line cannot be explained by the presence of an astrosphere around \aleo. 

\subsection{ Attempted detection of the warm/hot gas interface \label{sec:interface}} 
 It follows from the foregoing section that 
the nearest portion of the line  of sight from the Sun to \aleo\ has about 5\,pc of warm diffuse gas, and about 19\,pc of hot plasma. 
This implies that the line of sight crosses at least once the outer edge of the cloud.
 In the contact zone between the cloud warm gas and the hot gas two fundamental kinds of interactions may occur:  One is a conduction front, which, depending on the age of the interface, may be a conduit for either evaporation of the warm gas into the hot gas or the condensation of the hot gas onto the cooler material
\citep{Ballet1986,Boehringer.Hartquist1987, Slavin1989, 
Borkowski1990}. 
 The second kind of interaction is a turbulent mixing layer  that arises when the cool medium is moving relative to the hot gas and, as a result of instabilities, becomes entrained and mixed into the hot, more diffuse gas. Models for these  turbulent mixing layers were initially proposed by \cite{Begelman.Fabian1990} and further developed
by \cite{Slavin1993} and \cite{Kwak.Shelton2010}.  
Both types of interaction create regions of intermediate temperature that contain highly ionized metals, and all models predict substantial column densities for \CIV, \NV, \SiIV\ and \OVI. 

Our STIS spectrum of \aleo\ shows no evidence for any of the high ion species -- see the upper limits in Table~\ref{tab:results}. 
\cite{Snowden2015} also report the absence of a manifestation of an interface between the cold LLCC gas and the hot Local Bubble gas since  the X-ray data show no limb brightening at the edge of the LLCC (although they concede that such emission could occur at energies below the Rosat 1/4~keV passband).

Thermal conduction between the hot and warm media occurs primarily along magnetic field lines, and the electron heat flux is governed by the temperature gradient along such lines \citep{Balbus1986}.  \cite{Borkowski1990} showed that in a magnetized thermal conduction front, the predicted ion column densities vary with time, but more importantly, they can change by more than an order of magnitude depending on the inclination of the magnetic field relative to the front in the hot gas  $\theta_{h0}$.  We now address the issue of the possible influence of a magnetic field on the hot to warm interface in front of \aleo.

\cite{Zirnstein2016} found that the magnetic field at about 1000~AU from the Sun was oriented toward 227.28\degree$\pm$0.69\degree, 34.62\degree$\pm$0.45\degree\ in ecliptic
longitude and latitude. 
We convert this to $\ell\,\sim\, 26.1\degree$ and $b\,\sim\, 49.5\degree$ in Galactic coordinates. This means that in the direction to \aleo, 
the magnetic field makes an angle of 79\,\degree\  with respect to the direction of alpha Leo, i.e. it is almost perpendicular to the line of sight.

We have no explicit knowledge about the orientation of the interface's normal relative to our sight line, but if the angle between them is not very large, the field direction may be nearly parallel to the front. Moreover, this field is strongly coupled to the partly ionized warm medium and thus is probably influenced by forces on the gas near the front. It therefore follows that the field may be strengthened and pushed into closer alignment with the surface of the front if it is compressed so that it can brace the LIC to withstand the extra pressure from the hot medium (recall that the thermal pressure of the hot medium is probably larger than that of the LIC).  This configuration could steepen the temperature gradient across the front.

Fig~6 of \cite{Borkowski1990} shows that 
for $\theta_{h0}$ = 0\degree\ the column density predictions for \CIV\  (in log from 12.6 to 12.05) 
are above our detection limits 
 for a front younger than $10^6\,$yr, however when $\theta_{h0}$ = 60\degree\ or 
 85\degree\ the predicted column densities are below our detection limits for all ions. Therefore in the case where the hot gas magnetic field runs almost parallel to the cloud edge, which may be the case for the direction to \aleo, the interface high ion column densities could be reduced to the point that they are not detectable by our observations.

The \HI\ profile however indicates that an extra absorption is present that cannot be reconciled with our analysis of the low-ionization species (Section~\ref{sec:HI}). The contributions from hot gas   have already been invoked to interpret Lyman~$\alpha$ profiles  in low-column lines of sight, such as
$\epsilon$ CMa (Gry et al 1995), Sirius A \citep{Hebrard1999} and the astrospheres stars \citep{Wood2005}.

Let us first estimate the contribution of the Hot-Bubble gas to the profile. If this gas along the line of sight has the characteristics described by \cite{Snowden2014} over the path length discussed in Section~\ref{sec:filling} ($\sim 19$\,pc),  we estimate a total H column density of 2.7\,$10^{17} {\rm cm}^{-2}$. Under collisional ionization equilibrium (CIE) conditions, the neutral fraction expected in a gas at T$\sim 1.2\,10^6$ is equal to 1.9\,$10^{-7}$ \citep{Gnat.Sternberg2007} (calculated from the on-line tool provided by O.~Gnat) , yielding a neutral H column density of 5.1 $10^{10} {\rm cm}^{-2}$. At this temperature the $b$-value is 140\,\kms, which results in a profile width of FWHM = 0.95\AA\ that is compatible with the observed profile width (see Fig~\ref{fig:HI}), but the column density that we computed produces  a central optical depth of only  $2.8\, 10^{-4}$. The neutral hydrogen present amid the hot gas of the Local Bubble should therefore be undetectable in the Lyman $\alpha$ profile.

Alternatively, we may consider that the extra \HI\ absorption arises from the interface  between the cloud and the hot gas. At a typical interface temperature of 2\,$10^5$K, the  \HI\ column density  needed  to produce the missing absorption in the Lyman profile  is 

$N$(\HI) = 2.1\,$10^{15}  {\rm cm}^{-2}$ (shown in Fig~\ref{fig:HI}). If the gas were in CIE, this would imply
a total column density of 3.3\,$10^{20} {\rm cm}^{-2}$, 
which is unrealistically high  in view of the total column density of the cloud itself.
A more promising approach may be to consider the possibility that in the interface between the warm and hot phases, the collisional ionization of the hydrogen lags behind that expected for CIE as the temperature increases within the evaporation flow \citep{Ballet1986}.  In such a circumstance, the required total amount of hydrogen would be considerably less. We are not aware of explicit calculations discussed in the literature for the non-equilibrium expectations for $N$(H~I) in such flows at the level of thermal pressures that apply to our case.  It might be the case that the extra H~I absorption at large displacements from the line core could be a more sensitive indicator of an interface than any evidence provided by the highly ionized metal species. 
\subsection{Cloud ionization} 
 The neutral hydrogen density of $n({\rm H^0})$=0.20\,(+0.08,$-$0.1) ${\rm cm}^{-3}$ and the ionized hydrogen density of 
$n({\rm H^+})$=0.10(+0.02,$-$0.03) ${\rm cm}^{-3}$ lead to an ionization fraction of one third, $\chi$ = 0.33(+0.09,$-$0.06).

A notable outcome of our ionization model  is that the shielding column densities are 
substantially lower than half of $N({\rm H^0})$. This can plausibly be explained by the geometry of the LIC, which we propose is thinner in the  Canis Majoris direction. Supporting this idea is the fact that towards the star $\epsilon$~CMa, a powerful source of ionization of the local medium, the total \OI\ and \NI\ column densities are only 25\% of those towards \aleo\ \citep{Gry.Jenkins2001}. From this we infer a total neutral gas column density of $N$(\HI)$\sim5\,10^{17}$\,cm$^{-2}$ toward the ionizing source $\epsilon$ CMa, including about $\sim3.5\,10^{17}$\,cm$^{-2}$ for the LIC alone. Since \aleo\ is only 61\degree\ from $\epsilon$ CMa, and since the neutral gas subtends only a short distance, it is not difficult to imagine that the line of sight to \aleo\ runs  not far from the border on the exposed side of the LIC for most of 
 the cloud length, explaining the low shielding column density.

Table~\ref{tbl:ion_fractions} indicates for our preferred ionization condition how 
different elements are distributed among different ionization levels that are 
relevant.  The elements Si, S, and Fe have ion fractions near unity for their preferred 
stages of ionization, and this outcome holds for all acceptable ionization conditions.  
The elements C, N, O, Mg, and Ar have some concentrations in higher levels, and 
these fractions can vary within our most acceptable combinations of $n(e)$ and 
$n({\rm H_{tot}})$.
\begin{table}
\caption{Logarithms of the predicted ion 
fractions$^{a}$\label{tbl:ion_fractions}}
\begin{tabular}{
l	
c	
c	
c	
c	
}

 & {\footnotesize Log(X$^0$/X$_{\rm tot}$)}& 
{\footnotesize Log(X$^+$/X$_{\rm tot}$)} &{\footnotesize Log(X$^{++}$/X$_{\rm tot})$}& 
{\footnotesize Log(X$^{3+}$/X$_{\rm tot}$)}\\
               \hline 
{\footnotesize H}&$ -0.17$&$-0.48$&&      \\
{\footnotesize He}&$-0.28$&$-0.33$&$-1.75$&            \\
{\footnotesize C}& &$-0.15$&$-1.46$&$-6.02$\\
{\footnotesize N}&$-0.22$&$-0.40$&$-3.54$& \\
{\footnotesize O}&$-0.15$&$-0.52$&$-3.77$& \\
{\footnotesize Mg}& &$-0.16$&$-0.52$&$-5.56$\\
{\footnotesize Si}& &0.00&$-4.60$&$-8.36$\\
{\footnotesize S}& &$-0.05$&$-1.00$&$-5.33$\\
{\footnotesize Ar}&$-0.81$&$-0.07$&$-4.29$& \\
{\footnotesize Fe}& &0.00&$-3.03$&$-6.43$\\
               \hline

\end{tabular}
{\it $^a$} Applies to our most favored values $n(e)=0.11\,{\rm cm}^{-3}$ 
and $n({\rm H_{tot}})=0.3\,{\rm cm}^{-3}$.
\end{table}

Note that  He is always more ionized than H. The neutral fractions for He range 
from 0.42 to 0.57 while those for H range from 0.58 to 0.73.  Our calculated ratio $n({\rm He^0})/n({\rm H^0})$ varies from 
0.064 for $[n(e),n({\rm H_{tot}})]=[0.08,0.25]$ to 0.093 for $[n(e),n({\rm 
H_{tot}})]=[0.135,0.30]$.   These values for the neutral helium to hydrogen ratios 
are consistent with the findings for various local lines of sight considered by \cite{Dupuis1995} and \cite{Barstow1997}.

Our computed outcomes for the abundances of highly ionized species are lower than what was calculated by \cite{Slavin.Frisch2008}(model 26). This difference may be explained in part by their using a higher soft X-ray radiation field since they did not reduce the soft X-ray flux estimate for the contamination by a
contribution from the solar wind charge exchange.
\subsection{Depletion \label{sec:depletions}}
As we indicated in Section~\ref{sec:col_dens_other}, we find a substantial metal depletion in the line of sight toward \aleo, consistent for all elements with a line of sight depletion strength F$_*\,=\,0.6$ (following the scale established by Jenkins, 2009), except for magnesium which is even more depleted. 
The fact that depletions are present in our environment comes as no surprise, as 
this behavior has been noted elsewhere in the local ISM \citep{Wood2002,Slavin.Frisch2007, Frisch.Redfield.Slavin2011, Gry.Jenkins2014, Kimura2003, Kimura2015}.  
The fact that magnesium is over-depleted relative to other elements was already noticed by GJ14 for the bulk of sight-lines through the LIC. This may be relevant  to a particular change in the composition of dust in the solar neighborhood.

Table~\ref{tbl:depletions} lists the measured abundances of all observed species relative to the total hydrogen, after corrections for unseen ionization stages using the ionization fractions calculated in the model and given in Table~\ref{tbl:ion_fractions}. 
In calculating the depletions, we maintain consistency with \cite{Jenkins2009} by using as a comparison the recommended proto-solar abundance values by \cite{Lodders2003}, i.e. the  values of $A({\rm El})_0$ listed in her Table~2. The newer abundances by \cite{Lodders.Palme2009} differ by only $\pm$0.02 dex from those we used, except for Ar (-0.07 dex) and S (-0.05 dex).

For most studies of interstellar abundances, measurements of $N$(\AlII) are difficult because 
the \AlII\ absorption feature at 1671\,\AA\ is usually highly saturated.
However, observations over short paths in low density media avoid this problem and provide special opportunities to measure the depletion of aluminum in the gas phase. 
For the spectrum of \aleo,  our measurement of $N$(\AlII) from the unsaturated absorption feature and the upper limit on $N$(\AlIII), combined  with our estimate of $N({\rm H_{tot}}$),
yield an aluminum abundance  of log($N$(Al)/$N$(H))\,=\,$-7.00\,\pm0.18$. Compared to the solar abundance of aluminum [log(Al/H)$_{\odot} \,=\,-5.46 \pm0.02$ from \cite{Lodders2003}, we derive a depletion of [Al/H]\,=\,-1.54\,$\pm$0.18. In the line of sight to \aleo\ aluminum is significantly depleted in the gas phase with a depletion value similar to that of iron ($-1.60\,$dex).

It is noteworthy that our depletions for certain elements seem to be larger than typical values in the local region.  For instance, we find that $\log N({\rm Si~II})/N({\rm O~I})=-1.95$ for \aleo\ is surpassed by only 4 out of 32 sight lines studied by \cite{Redfield.Linsky2004a}.  Likewise, our determination $\log N({\rm Al~II})/N({\rm O~I})=-3.61$ represents a depletion of Al that is matched or exceeded by only 2 out of 17 cases examined in this same investigation, not counting upper limits that were weaker (i.e., larger) than our result.
GJ14 had noted the existence of an abundance gradient for Fe and Mg within the LIC, with their depletions increasing along the direction of motion from the rear to the front of the cloud. Although the angle between the direction of motion and the direction of \aleo\  is 76\degree, placing this sight line in the middle zone with intermediate depletion, the Mg and Fe depletion  values found toward \aleo\ are consistent with the highest depletions\footnote{The GJ14 depletion values for Fe should be increased by $-0.17$ to account for the ionization of H. No such correction is needed for Mg, since the ionization fraction of \MgII\ is similar to that of \HI\ in our model.}, found by GJ14 in the head of the cloud. 
In the direction toward Capella (20\degree\ from the head of the cloud) \cite{Wood2002} have derived depletion values in the LIC for all our elements but argon and sulfur. All values are compatible with our findings and the agreement is excellent for Mg, as well as for Al and Si.

 Our derived depletion strength F$_*\,=\,0.6$ (Section~\ref{sec:col_dens_other}) is higher than any of the synthetic F$_*$ values\footnote{A synthetic F$_*$ is a depletion outcome that is based solely on the relative abundances of various elements, without reference to the amount of hydrogen present.}  found by \cite{Jenkins2009} for WD stars in the Local Bubble with an average value of F$_{*sync}\,=\,0.15\pm 0.23$. These stars are generally more distant than \aleo, but they give the general trend for sight lines that penetrate other low density clouds in the Local Bubble. However,  \cite{Jenkins2009} notes that six of the stars have F$_{*sync}\,\geq 0.3$, indicating that moderately strong depletions do exist in our general vicinity.

We note that both oxygen and argon appear to be undepleted in the gas phase.  By contrast, sulfur shows a depletion by almost a factor of two. 
 Sulfur has often been considered to be undepleted, and as such has been used in many studies as a standard for zero depletion. \cite{Jenkins2009} presented some evidence for S depletion but noted that S is a troublesome element because  N(\SII) usually cannot be reliably measured except for cases where column densities are low.  Furthermore because of its high ionization potential (23 eV), some of the \SII\ can arise from the  H II region around the target star. The  line of sight toward \aleo,  where the \SII\ lines are not saturated and the total H column density is known, provides an important indication that sulfur can indeed be depleted in the ISM. 

Carbon may be slightly overabundant, although its overabundance is not significant after considering the error bars on $N$(C~II). An overabundance of carbon in the local ISM had already been proposed by \cite{Slavin.Frisch2006}. We find towards \aleo\ a gas-phase abundance close to their lower limit of 330\,ppm. We confirm here their finding that the carbon to sulfur ratio is higher than solar, with C/S = 34 (+13,-6), to be compared with the solar ratio of 16. 
However towards \aleo\ this indicates  sulfur depletion rather than an  overabundance of carbon.

\begin{table}[h]
\caption{Logarithms of elemental abundances and depletion values.  \label{tbl:depletions}}
\begin{tabular}{
l	
c	
c	
}
 & {\footnotesize abundance Log(X$^{a}_{\rm tot}$/H$^{b}_{\rm tot})_{\rm obs}$ }& {\footnotesize depletion$^{c}$ [X/H]}\\
               \hline 
{\footnotesize C}&$-3.48$ &$+0.06$\\
{\footnotesize N}&$-4.21$&$-0.11$ \\
{\footnotesize O}&$-3.24$&$-0.00$ \\
{\footnotesize Mg}&$-5.48$&$-1.10$\\
{\footnotesize Al}&$-7.00$&$-1.54$\\
{\footnotesize Si}&$-5.34$&$-0.95$\\
{\footnotesize S}& $-5.01$&$-0.27$\\
{\footnotesize Ar}&$-5.37$&$+0.01$ \\
{\footnotesize Fe}&$-6.06$&$-1.60$\\
               \hline 
\end{tabular}

{\it $^{a}$ }Measured column density corrected for the ion fraction \\
{\it $^b$} Derived with  $N$(H$_{tot}$)=2.83\,$10^{18}$\cms. \\
{\it $^c$} Reference abundances: the protosolar values (current solar photospheric abundances +0.074 dex to correct for gravitational settling) taken from \cite{Lodders2003}.  These abundances are consistent with those used by \cite{Jenkins2009}.
\end{table}

\subsection{Secondary components \label{sec:second}}
\begin{figure}[h]
\includegraphics[width=1.0\columnwidth]{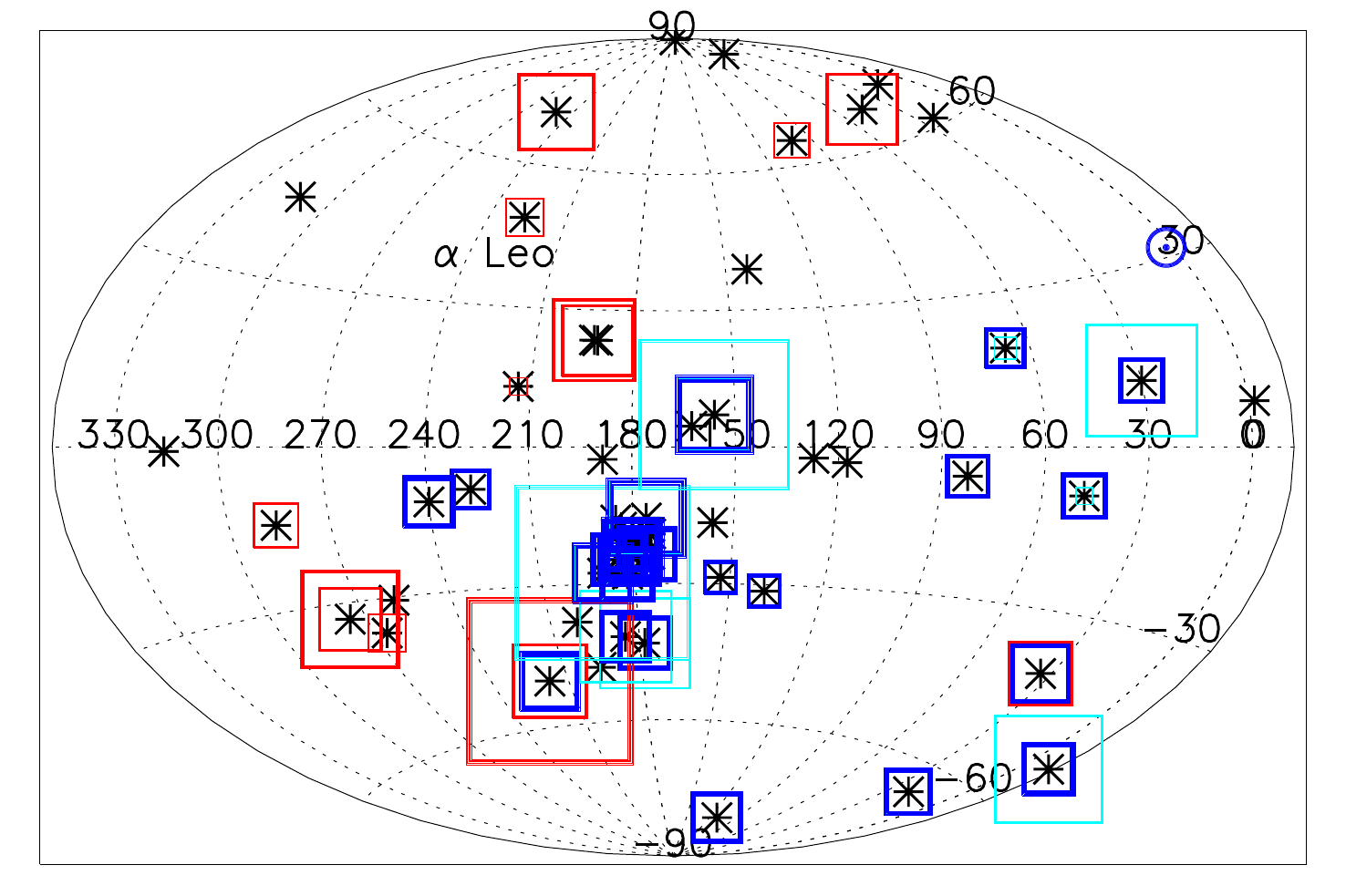} 
\caption{A map of secondary components showing their velocity shift relative to the LIC component. This plot is the same as Fig.~12 of \cite{Gry.Jenkins2014}, but with one correction and a few additions including \aleo. Black asterisks signal the position of all targets.   Blue/red squares mark secondary components with negative/positive shifts, and the size of the symbols scales with the amplitude of the shift.   Dark blue symbols indicate the  components having a uniform velocity shift of 
$-7.2\pm$1.5 ${\rm km~s}^{-1}$ relative to the LIC (i.e. the Cetus Ripple); light blue symbols represent the other blue-shifted components. 
The blue circle at l = 2.3\degree, b = 31.4\degree shows the "Warm Breeze" heliospheric inflow direction. 
\label{fig:second}}
\end{figure}

While our ionization analysis applies to all of the matter present in the line of sight, we recognize that the gas is distributed between two distinct velocity components separated by 5.6 \kms. The dominant component is the LIC with $\approx 75$\% of the total column density. As mentioned in Section~\ref{sec:kinematics}, the second component is at a positive shift relative to the LIC.  This behavior is consistent with that of other secondary components in this region of the sky, as shown in Fig.~\ref{fig:second}, which is an update of Fig.~12 of GJ14.

In GJ14 we had shown that a secondary component moving towards the cloud center with a relative velocity close to $-7\,$\kms\
was present in all sight-lines on half of the sky (shown in dark blue on Fig.~\ref{fig:second}). We called this component the Cetus Ripple in view of the rough location in the sky of its center, and we showed that all observed characteristics were consistent with a shock travelling in the LIC toward the cloud interior at a velocity of 20 to 26 \kms. On the other side of the sky, where \aleo\ is located, all secondary components have  a positive velocity shift relative to the LIC, meaning that they move away from it or that they move outwards if they are included in the cloud.

As shown in the last column of Table~\ref{tab:results} the proportion of gas at the LIC velocity versus the proportion of gas in the second component is very stable from species to species, which indicates similar conditions in both components. This makes it plausible that the second component is but an inclusion within the LIC.

This second component, along with other secondary components seen in the direction of the nearest stars, may be related to kinematic disturbances within the local cloud. They may represent
internal flows of matter with different velocities in the form of waves propagating in the local cloud matter. 

In this context, we may try to
 establish a connection with the IBEX ``Warm Breeze", a second component of neutral He discovered in the IBEX data, 
 with inflow Galactic coordinates $\ell = 2.3\degree, b = 31.4\degree$,
 with a different velocity, temperature and density than the main  component that corresponds to the LIC flowing into the heliosphere. \cite{Kubiak2014} who report its discovery  offer several possible interpretations involving heliospheric configurations, and they also propose that the Warm Breeze might be a flow of interstellar matter not thermalized with the LIC material, moving in the LIC with a relative velocity equal to  V$_{WB}\, -\,V_{LIC}$ =  13.6 \kms. 
   Although our secondary components are in principle more extended than the IBEX Warm Breeze (they 
represent a substantial  portion of the matter on most lines of sight even if the LIC is the dominant component), the Warm Breeze could be the local signature of an interstellar component of the same nature as the secondary components present in many stellar lines of sight.

\section{Implications for  the Local Leo Cold Cloud  (LLCC) \label{sec:LLCC}}
Results discussed in the previous sections show that the interstellar absorption features detected in the line of sight toward \aleo\ are fully consistent in both velocity and magnitude with what we  expect from the local ISM.
There is no evidence for the presence of  cold and dense gas of the type seen in the interior of the LLCC, and also no evidence for an excess of  warm diffuse gas. 

The good coincidence in velocity between the LIC in this direction and the LLCC  is however worth pointing out. 
By performing a least-square fit to the velocities measured in the main body of the LLCC, Peek et al (2011) predicted an LSR velocity of 3.19 \kms\ at the position of \aleo\ for the LLCC assuming that it is moving as a solid body. This velocity translates to a heliocentric velocity of 9.3 \kms, which is in excellent agreement with GJ14 prediction for the LIC velocity. 
The coincidence is even more compelling when we notice that it holds true along the long string of dense clouds \cite{Haud2010} that stretches across 80\degree\ in the sky  and that includes the LLCC. Along this string the GJ14 model for the deformed LIC  predicts LSR velocities from 0 \kms\ to 12.7 \kms\ from its northern location around ($\ell,b$)=(240\degree, +46\degree) to its south-western end around (180\degree,20\degree). This is in intriguingly good agreement with the color-coded cold matter velocities displayed all along the string in Figure 2-a of \cite{Haud2010}
  
 The weak \CaII\  absorption feature in the optical spectrum of \aleo\ studied in  detail by \cite{Peek2011}, is detected at an LSR velocity of 3.1\,\kms.   Since it is  identical to the LLCC velocity, \cite{Peek2011} claimed that the  Ca~II absorption feature arises from the outermost portions of the LLCC extending across the \aleo\ line of sight. 
This interpretation is however challenged by the coincidence of this feature with the properties of the LIC found by us and other investigators, as we discuss below.
 
 The LSR velocity quoted by \cite{Peek2011}, which we convert to a heliocentric velocity of +9.2\,\kms, coincides with that of our Component~1 and the GJ14 prediction of the LIC velocity. 
\cite{Frisch2002} refer to a Ca II spectrum taken with a resolution of about 1.3 \kms, 
and  with a signal-to-noise ratio of about 500. A fit to the normalized line profile as a single component yielded ({\it D. Welty, private communication}) 
b $\sim$ 3.6 $\pm$ 0.7 \kms, and $v_{hel}\,\sim 10.5 \pm 0.5\,\kms$, which are values that are consistent with our two-component model for \aleo, if expressed in just the context of a single component.
 Futhermore the derived column density is N(\CaII) $\sim$ 5.0 $\pm 0.8\, 10^9$ \cms, which is generally
typical of values found for the LIC \citep{Lallement.Bertin1992}.
For example, it is identical to the \CaII\ column densities 
listed for the LIC towards $\alpha$ Aql (5.1 pc, N(Ca II)$\sim7 \, 10^9$ \cms; \cite{Ferlet1986} or $\delta$ Cas (30.5 pc, N(Ca II)=5.0 $\pm 2.5\, 10^9$ \cms ; \cite{Lallement.Bertin1992}, which both show \MgII\ column densities very close to those we find for \aleo\ \cite{Lallement1996, Lallement.Ferlet1997}.  
 
 We conclude that the \CaII\ feature is consistent with being produced in the LIC. 
Now, if this feature is not attributed to the LLCC, there is no  evidence that the LLCC must be located in front of \aleo\ [as had been proposed  by \cite{Peek2011}].  It follows that the distance to \aleo\ (24 pc) no longer represents a reliable upper limit for the distance to the LLCC. The only firm upper limit to the distance of the cloud  therefore remains at 45 pc, as established by \cite{Meyer2006}.

This has an important implication regarding the understanding of the hot gas observed in the Local Bubble. 
 \cite{Snowden2015} showed that the soft X-ray foreground emission to the LLCC ``is only marginally consistent with the range of possible LHB plasma path lengths between the LLCC and the LIC''. This conclusion was drawn by adopting 24 pc as upper limit to the LLCC distance. Since we have now shown that this upper limit is no longer warranted, the constraint is much less stringent, and there is room for their required path length  D$^{LHB}=29 \pm 11 pc$ for the hot gas in front of the LLCC. 

With the observation of the hot gas foreground emission occurring over a length  of $29 \pm 11 pc$  \citep{Snowden2015} we obtain a very consistent picture where the warm, diffuse, partially neutral, gas extends out for a total of $3(+5,-1) pc$ complemented by $29 \pm 11 pc$ of hot gas, until it reaches the cold dense matter of the LLCC at a distance of 33.5 $\pm 11.3 \, pc$. This maximum distance derived from the  path length of the soft X-ray emission is now fully consistent with the LLCC distance upper limit derived from the absorption spectra of background stars.

Comparing the path length  derived for the warm diffuse gas observed in the \aleo\ line of sight (between 2 and 8 pc for the sum of our two components) to the  minimum distance of the LLCC of 11 pc \cite{Peek2011}, we could 
 conclude that it is likely that the two clouds are not physically contiguous, if the LIC is supposed to be a continuous medium.

Yet, the excellent 
 kinematical coincidences also lead us to put forward the hypothesis  that the LLCC, as well as the Haud ring, are part of a complex of gas that may somehow be physically related to the LIC.
 In the picture of the LIC being a continuous medium,  we recognize that 
the LLCC and the LIC seem to be separated by a long path of hot gas. However in the hypotheses of a 
patchy cloud, the low filling factor could also be interpreted as the presence of  voids within the volume of the LIC, the LIC presenting  a "Swiss cheese'' structure or  a filamentary structure, warm gas filaments alternating with voids filled with hot gas. In this case, the overall volume occupied by the diffuse local cloud would reach further out, out to the distance of the LLCC.  In this hypothesis the LIC  could spatially coincide with outer regions of the LLCC, or the LLCC could be a clump of cold dense gas formed inside the diffuse warm gas of the LIC. 

\section{Summary and conclusion  \label{conclusion}}
{
We have analyzed the ultraviolet absorption spectrum of the nearby star \aleo\ to characterize  the interstellar medium in its line of sight, and in particular the warm diffuse interstellar cloud surrounding the Sun.
\\
- The strongest of the two velocity components coincides with the predictions for the local interstellar cloud (LIC) detected in all directions around the Sun, according to the picture proposed by Gry.Jenkins (2014).
 The second, smaller, component exhibits a velocity shift of +5.4\,\kms\ relative to the main component, in agreement with shifts observed in other lines of sight in this region of the sky. 
 \\
- By performing profile fitting on the spectra, we derive column density intervals or upper limits for 18 atoms or ions (Table~\ref{tab:results}).
The LIC contains about 75 \% of the total matter in the line of sight and the distribution between both components does not vary significantly from element to element.
 \\
-  The temperature $T$ and the electron density $n(e)$ are derived by combining the measurements of the fine-structure level excitation of C~II and the ionization equilibrium  of Mg~I and Mg~II.
\\ 
- After estimating the local interstellar radiation field in the UV, extreme UV, and X-ray domains (Fig.~\ref{fig:ISRF}) and considering all possible ionizing and recombination processes, we study the ionization balance of all elements and create a model that describes consistently the partial ionization of the gas. Two free parameters in this model are (1) the amount of shielding of this radiation by neutral hydrogen and helium and (2) the volume density of hydrogen.
 \\
-  The total (neutral plus ionized) amount of hydrogen $N({\rm H_{tot}})$ is derived from the sum of the \NI\ and \NII\ column densities.
\\ 
 - The densities of neutral and ionized hydrogen, and hence $N({\rm H~I})$, result from our model for the photoionization of hydrogen and our knowledge of $n(e)$.
 \\
- The  ionization fractions of all elements follow from our model (Table~\ref{tbl:ion_fractions}), and their comparisons with the observations provide a measure of the depletion of metals in the gas phase. 
\\
-Table~\ref{tab:LIC} summarizes  the characteristics derived from this analysis for the warm interstellar matter in the \aleo\ sight line. 
 \begin{table}[h]
\caption{Characteristics of the warm interstellar gas in the line of sight of \aleo\ \label{tab:LIC}}
\begin{tabular}{ll}
\hline
$N({\rm H_{tot}})$ (\cms) & 2.83\,$^{+1.18}_{-0.69}\,10^{18}$\\[4pt]
 $N$(\HI) (\cms) & 1.9\,$^{+0.9}_{-0.6}$\,10$^{18}$\\[4pt]
$n({\rm H_{tot}})$ (cm$^{-3}$) & $0.30\,^{+0.10}_{-0.13}$\\[4pt]
$n$(\HI) (cm$^{-3}$)& $0.20\,^{+0.08}_{-0.10}$ \\[4pt]
$n(e)$ (cm$^{-3}$)&  0.11\,$^{+0.025}_{-0.03}$\\[4pt]
$T$ (K) & 6500\,$^{+750}_{-600}$\\[4pt]
Pressure log$(p/k)$ & 3.42\,$^{+0.12}_{-0.22}$\\[4pt]
Path length (pc)& 3\,$^{+5}_{-1}$\\[4pt]
ioniz. fraction $\chi$ & 0.33\,$^{+0.09}_{-0.06}$\\[4pt]
depletion strength$^a$ $F_*$ & 0.6\\
\hline
$^a$ {\footnotesize in the sense defined by Jenkins (2009)}
\end{tabular}
\end{table}
With the exception of  the temperatures and the electron densities that we derived for each of the velocity components, the  analysis has been done for the combination of the two components. 
Nevertheless  since we have shown that the conditions in both components  are likely to be identical, the densities and the ionization ratios found for them are likely to apply to just the LIC alone. 
\\   
-  From the neutral hydrogen column and volume densities we derive a filling factor of only 13\%  for the warm gas, the remaining space is probably filled with hot, soft-X-ray emitting gas, in agreement with measurements of the diffuse soft X-ray background radiation.
 \\
-  We do not detect any  absorption features in the highly ionized species that could be produced in the interfaces between the warm clouds and the surrounding hot gas, possibly because of the reduction of thermal conduction due to  the alignment of the magnetic field with the surface of the conduction front. On the other hand,  an extra component of hot neutral hydrogen, required to fit the  Lyman $\alpha$ absorption feature, may turn out to be the best  interface signature.
\\  
- This sight-line is particularly interesting because it is close to a nearby cold cloud called the Local Leo Cold Cloud (LLCC) discovered by \cite{Meyer2006}. We show that the Ca~II absorption that had been invoked to set a new upper limit of 24 pc for the distance of LLCC  is actually fully consistent with just the LIC absorption. 
Therefore,  the LLCC upper limit remains 42 pc and this reconciles the LLCC distance with the estimated path length of the measured X-ray emission from the foreground   hot gas in the Local Bubble, 
which places the LLCC at a distance of 33.5 $\pm 11.3 \, pc$. 
\\
- We note the remarkable velocity coincidence between the LIC and the LLCC. We also note that in the hypothesis that the LIC has a patchy structure, with diffuse warm gas alternating with hot gas along the line of sight, the LIC could reach as far as the distance of the LLCC and the two clouds may be physically related.

\begin{acknowledgements}
Based on observations made with the NASA/ESA Hubble Space Telescope that were obtained for the ASTRAL Treasury program during the Cycle~21 observing session. Support for the research by EBJ was provided by NASA through a grant HST-GO-13346.15-A to Princeton University from the Space Telescope Science Institute (STScI), which is operated by the Association of Universities for Research in Astronomy, Inc. under the NASA contract NAS 5-26555.  The {\it Copernicus\/} data were obtained from the Mikulski Archive for Space Telescopes (MAST) at STScI, supported by the NASA Office of Space Science via grant NNX09AF08G and by other grants and contracts.     EBJ is grateful to Aix-Marseille Universit\'e for  providing a visitor grant to support his 1-month visit to LAM where this research was initiated.  We are thankful to T.~Ayres for his dedication and skill in organizing and programming the observations that made up the Hot Stars episode of the ASTRAL program.
\end{acknowledgements}

\bibliographystyle{aa}

\bibliography{biblio-2016}

\begin{thebibliography}{128}
\expandafter\ifx\csname natexlab\endcsname\relax\def\natexlab#1{#1}\fi

\bibitem[{{Absil} {et~al.}(2011){Absil}, {Le Bouquin}, {Berger}, {Lagrange},
  {Chauvin}, {Lazareff}, {Zins}, {Haguenauer}, {Jocou}, {Kern}, {Millan-Gabet},
  {Rochat}, \& {Traub}}]{Absil2011}
{Absil}, O., {Le Bouquin}, J.-B., {Berger}, J.-P., {et~al.} 2011, \aap, 535,
  A68

\bibitem[{{Aldrovandi} \& {Pequignot}(1973)}]{Aldrovandi.Pequignot1973}
{Aldrovandi}, S.~M.~V. \& {Pequignot}, D. 1973, \aap, 25, 137

\bibitem[{{Allan} {et~al.}(1988){Allan}, {Clegg}, {Dickinson}, \&
  {Flower}}]{Allan1988}
{Allan}, R.~J., {Clegg}, R.~E.~S., {Dickinson}, A.~S., \& {Flower}, D.~R. 1988,
  \mnras, 235, 1245

\bibitem[{{Altun} {et~al.}(2006){Altun}, {Yumak}, {Badnell}, {Loch}, \&
  {Pindzola}}]{Altun2006}
{Altun}, Z., {Yumak}, A., {Badnell}, N.~R., {Loch}, S.~D., \& {Pindzola}, M.~S.
  2006, \aap, 447, 1165

\bibitem[{{Badnell}(2006)}]{Badnell2006}
{Badnell}, N.~R. 2006, \apjs, 167, 334

\bibitem[{{Bahcall} \& {Wolf}(1968)}]{Bahcall.Wolf1968}
{Bahcall}, J.~N. \& {Wolf}, R.~A. 1968, \apj, 152, 701

\bibitem[{{Baker} \& {Menzel}(1938)}]{Baker.Menzel1938}
{Baker}, J.~G. \& {Menzel}, D.~H. 1938, \apj, 88, 52

\bibitem[{{Balbus}(1986)}]{Balbus1986}
{Balbus}, S.~A. 1986, \apj, 304, 787

\bibitem[{{Ballet} {et~al.}(1986){Ballet}, {Arnaud}, \&
  {Rothenflug}}]{Ballet1986}
{Ballet}, J., {Arnaud}, M., \& {Rothenflug}, R. 1986, \aap, 161, 12

\bibitem[{{Barinovs} {et~al.}(2005){Barinovs}, {van Hemert}, {Krems}, \&
  {Dalgarno}}]{Barinovs2005}
{Barinovs}, {\u G}., {van Hemert}, M.~C., {Krems}, R., \& {Dalgarno}, A. 2005,
  \apj, 620, 537

\bibitem[{{Barstow} {et~al.}(2014){Barstow}, {Barstow}, {Casewell}, {Holberg},
  \& {Hubeny}}]{Barstow2014}
{Barstow}, M.~A., {Barstow}, J.~K., {Casewell}, S.~L., {Holberg}, J.~B., \&
  {Hubeny}, I. 2014, \mnras, 440, 1607

\bibitem[{{Barstow} {et~al.}(2010){Barstow}, {Boyce}, {Welsh}, {Lallement},
  {Barstow}, {Forbes}, \& {Preval}}]{Barstow2010}
{Barstow}, M.~A., {Boyce}, D.~D., {Welsh}, B.~Y., {et~al.} 2010, \apj, 723,
  1762

\bibitem[{{Barstow} {et~al.}(1997){Barstow}, {Dobbie}, {Holberg}, {Hubeny}, \&
  {Lanz}}]{Barstow1997}
{Barstow}, M.~A., {Dobbie}, P.~D., {Holberg}, J.~B., {Hubeny}, I., \& {Lanz},
  T. 1997, \mnras, 286, 58

\bibitem[{{Barstow} {et~al.}(2003){Barstow}, {Good}, {Holberg}, {Hubeny},
  {Bannister}, {Bruhweiler}, {Burleigh}, \& {Napiwotzki}}]{Barstow2003}
{Barstow}, M.~A., {Good}, S.~A., {Holberg}, J.~B., {et~al.} 2003, \mnras, 341,
  870

\bibitem[{{Begelman} \& {Fabian}(1990)}]{Begelman.Fabian1990}
{Begelman}, M.~C. \& {Fabian}, A.~C. 1990, \mnras, 244, 26P

\bibitem[{{Begum} {et~al.}(2010){Begum}, {Stanimirovi{\'c}}, {Peek},
  {Ballering}, {Heiles}, {Douglas}, {Putman}, {Gibson}, {Grcevich}, {Korpela},
  {Lee}, {Saul}, \& {Gallagher}}]{Begum2010}
{Begum}, A., {Stanimirovi{\'c}}, S., {Peek}, J.~E., {et~al.} 2010, \apj, 722,
  395

\bibitem[{{Boehringer} \& {Hartquist}(1987)}]{Boehringer.Hartquist1987}
{Boehringer}, H. \& {Hartquist}, T.~W. 1987, \mnras, 228, 915

\bibitem[{{Bohlin}(1975)}]{Bohlin1975}
{Bohlin}, R.~C. 1975, \apj, 200, 402

\bibitem[{{Borkowski} {et~al.}(1990){Borkowski}, {Balbus}, \&
  {Fristrom}}]{Borkowski1990}
{Borkowski}, K.~J., {Balbus}, S.~A., \& {Fristrom}, C.~C. 1990, \apj, 355, 501

\bibitem[{{Burlaga} \& {Ness}(2014)}]{Burlaga.Ness2014}
{Burlaga}, L.~F. \& {Ness}, N.~F. 2014, \apjl, 795, L19

\bibitem[{{Butler} \& {Dalgarno}(1980)}]{Butler.Dalgarno1980}
{Butler}, S.~E. \& {Dalgarno}, A. 1980, \apj, 241, 838

\bibitem[{{Butler} {et~al.}(1980){Butler}, {Heil}, \& {Dalgarno}}]{Butler1980}
{Butler}, S.~E., {Heil}, T.~G., \& {Dalgarno}, A. 1980, \apj, 241, 442

\bibitem[{{Cowie} \& {McKee}(1977)}]{Cowie.McKee1977}
{Cowie}, L.~L. \& {McKee}, C.~F. 1977, \apj, 211, 135

\bibitem[{{Craig} {et~al.}(1997){Craig}, {Abbott}, {Finley}, {Jessop},
  {Howell}, {Mathioudakis}, {Sommers}, {Vallerga}, \& {Malina}}]{Craig1997}
{Craig}, N., {Abbott}, M., {Finley}, D., {et~al.} 1997, \apjs, 113, 131

\bibitem[{{Dalton} \& {Balbus}(1993)}]{Dalton.Balbus1993}
{Dalton}, W.~W. \& {Balbus}, S.~A. 1993, \apj, 404, 625

\bibitem[{{Draine}(2011)}]{Draine2011}
{Draine}, B.~T. 2011, {Physics of the Interstellar and Intergalactic Medium}
  (Princeton University Press)

\bibitem[{{Dupuis} {et~al.}(1995){Dupuis}, {Vennes}, {Bowyer}, {Pradhan}, \&
  {Thejll}}]{Dupuis1995}
{Dupuis}, J., {Vennes}, S., {Bowyer}, S., {Pradhan}, A.~K., \& {Thejll}, P.
  1995, \apj, 455, 574

\bibitem[{{Ferlet} {et~al.}(1986){Ferlet}, {Vidal-Madjar}, \&
  {Lallement}}]{Ferlet1986}
{Ferlet}, R., {Vidal-Madjar}, A., \& {Lallement}, R. 1986, \aap, 163, 204

\bibitem[{{Frisch} {et~al.}(2002){Frisch}, {Grodnicki}, \&
  {Welty}}]{Frisch2002}
{Frisch}, P.~C., {Grodnicki}, L., \& {Welty}, D.~E. 2002, \apj, 574, 834

\bibitem[{{Frisch} {et~al.}(2011){Frisch}, {Redfield}, \&
  {Slavin}}]{Frisch.Redfield.Slavin2011}
{Frisch}, P.~C., {Redfield}, S., \& {Slavin}, J.~D. 2011, \araa, 49, 237

\bibitem[{{Galeazzi} {et~al.}(2014){Galeazzi}, {Chiao}, {Collier}, {Cravens},
  {Koutroumpa}, {Kuntz}, {Lallement}, {Lepri}, {McCammon}, {Morgan}, {Porter},
  {Robertson}, {Snowden}, {Thomas}, {Uprety}, {Ursino}, \&
  {Walsh}}]{Galeazzi2014}
{Galeazzi}, M., {Chiao}, M., {Collier}, M.~R., {et~al.} 2014, \nat, 512, 171

\bibitem[{{Geary}(1930)}]{Geary1930}
{Geary}, R.~C. 1930, Journal of the Royal Statistical Society, 93, 442

\bibitem[{{Gies} {et~al.}(2008){Gies}, {Dieterich}, {Richardson}, {Riedel},
  {B.~L.~Team}, {McAlister}, {Bagnuolo}, {Grundstrom}, {{\v S}tefl},
  {Rivinius}, \& {Baade}}]{Gies2008}
{Gies}, D.~R., {Dieterich}, S., {Richardson}, N.~D., {et~al.} 2008, \apjl, 682,
  L117

\bibitem[{{Gnat} \& {Sternberg}(2007)}]{Gnat.Sternberg2007}
{Gnat}, O. \& {Sternberg}, A. 2007, \apjs, 168, 213

\bibitem[{{Gondhalekar} {et~al.}(1980){Gondhalekar}, {Phillips}, \&
  {Wilson}}]{Gondhalekar1980}
{Gondhalekar}, P.~M., {Phillips}, A.~P., \& {Wilson}, R. 1980, \aap, 85, 272

\bibitem[{{Gry} \& {Jenkins}(2001)}]{Gry.Jenkins2001}
{Gry}, C. \& {Jenkins}, E.~B. 2001, \aap, 367, 617

\bibitem[{{Gry} \& {Jenkins}(2014)}]{Gry.Jenkins2014}
{Gry}, C. \& {Jenkins}, E.~B. 2014, \aap, 567, A58

\bibitem[{{Haud}(2010)}]{Haud2010}
{Haud}, U. 2010, \aap, 514, A27

\bibitem[{{H{\'e}brard} {et~al.}(2002){H{\'e}brard}, {Lemoine}, {Vidal-Madjar},
  {D{\'e}sert}, {Lecavelier des {\'E}tangs}, {Ferlet}, {Wood}, {Linsky},
  {Kruk}, {Chayer}, {Lacour}, {Blair}, {Friedman}, {Moos}, {Sembach},
  {Sonneborn}, {Oegerle}, \& {Jenkins}}]{Hebrard2002}
{H{\'e}brard}, G., {Lemoine}, M., {Vidal-Madjar}, A., {et~al.} 2002, \apjs,
  140, 103

\bibitem[{{H{\'e}brard} {et~al.}(1999){H{\'e}brard}, {Mallouris}, {Ferlet},
  {Koester}, {Lemoine}, {Vidal-Madjar}, \& {York}}]{Hebrard1999}
{H{\'e}brard}, G., {Mallouris}, C., {Ferlet}, R., {et~al.} 1999, \aap, 350, 643

\bibitem[{{Heiles} \& {Troland}(2003)}]{Heiles.Troland2003}
{Heiles}, C. \& {Troland}, T.~H. 2003, \apj, 586, 1067

\bibitem[{{Indebetouw} \& {Shull}(2004{\natexlab{a}})}]{Indebetouw.Shull2004a}
{Indebetouw}, R. \& {Shull}, J.~M. 2004{\natexlab{a}}, \apj, 605, 205

\bibitem[{{Indebetouw} \& {Shull}(2004{\natexlab{b}})}]{Indebetouw.Shull2004b}
{Indebetouw}, R. \& {Shull}, J.~M. 2004{\natexlab{b}}, \apj, 607, 309

\bibitem[{{Indriolo} \& {McCall}(2012)}]{Indriolo.McCall2012}
{Indriolo}, N. \& {McCall}, B.~J. 2012, \apj, 745, 91

\bibitem[{{Indriolo} {et~al.}(2015){Indriolo}, {Neufeld}, {Gerin}, {Schilke},
  {Benz}, {Winkel}, {Menten}, {Chambers}, {Black}, {Bruderer}, {Falgarone},
  {Godard}, {Goicoechea}, {Gupta}, {Lis}, {Ossenkopf}, {Persson},
  {Sonnentrucker}, {van der Tak}, {van Dishoeck}, {Wolfire}, \&
  {Wyrowski}}]{Indriolo2015}
{Indriolo}, N., {Neufeld}, D.~A., {Gerin}, M., {et~al.} 2015, \apj, 800, 40

\bibitem[{{Jenkins}(2002)}]{Jenkins2002}
{Jenkins}, E.~B. 2002, \apj, 580, 938

\bibitem[{{Jenkins}(2009)}]{Jenkins2009}
{Jenkins}, E.~B. 2009, \apj, 700, 1299

\bibitem[{{Jenkins}(2013)}]{Jenkins2013}
{Jenkins}, E.~B. 2013, \apj, 764, 25

\bibitem[{{Jenkins} {et~al.}(2000){Jenkins}, {Oegerle}, {Gry}, {Vallerga},
  {Sembach}, {Shelton}, {Ferlet}, {Vidal-Madjar}, {York}, {Linsky}, {Roth},
  {Dupree}, \& {Edelstein}}]{Jenkins2000}
{Jenkins}, E.~B., {Oegerle}, W.~R., {Gry}, C., {et~al.} 2000, \apjl, 538, L81

\bibitem[{{Jenkins} \& {Tripp}(2011)}]{Jenkins.Tripp2011}
{Jenkins}, E.~B. \& {Tripp}, T.~M. 2011, \apj, 734, 65

\bibitem[{{Kimura}(2015)}]{Kimura2015}
{Kimura}, H. 2015, \mnras, 449, 2250

\bibitem[{{Kimura} {et~al.}(2003){Kimura}, {Mann}, \&
  {Jessberger}}]{Kimura2003}
{Kimura}, H., {Mann}, I., \& {Jessberger}, E.~K. 2003, \apj, 582, 846

\bibitem[{{Kingdon} \& {Ferland}(1996)}]{Kingdon.Ferland1996}
{Kingdon}, J.~B. \& {Ferland}, G.~J. 1996, \apjs, 106, 205

\bibitem[{{Kubiak} {et~al.}(2014){Kubiak}, {Bzowski}, {Sok{\'o}{\l}},
  {Swaczyna}, {Grzedzielski}, {Alexashov}, {Izmodenov}, {M{\"o}bius},
  {Leonard}, {Fuselier}, {Wurz}, \& {McComas}}]{Kubiak2014}
{Kubiak}, M.~A., {Bzowski}, M., {Sok{\'o}{\l}}, J.~M., {et~al.} 2014, \apjs,
  213, 29

\bibitem[{{Kuntz} \& {Snowden}(2000)}]{Kuntz.Snowden2000}
{Kuntz}, K.~D. \& {Snowden}, S.~L. 2000, \apj, 543, 195

\bibitem[{{Kwak} \& {Shelton}(2010)}]{Kwak.Shelton2010}
{Kwak}, K. \& {Shelton}, R.~L. 2010, \apj, 719, 523

\bibitem[{{Lallement} \& {Bertin}(1992)}]{Lallement.Bertin1992}
{Lallement}, R. \& {Bertin}, P. 1992, \aap, 266, 479

\bibitem[{{Lallement} \& {Ferlet}(1997)}]{Lallement.Ferlet1997}
{Lallement}, R. \& {Ferlet}, R. 1997, \aap, 324, 1105

\bibitem[{{Lallement} {et~al.}(1995){Lallement}, {Ferlet}, {Lagrange},
  {Lemoine}, \& {Vidal-Madjar}}]{Lallement1995}
{Lallement}, R., {Ferlet}, R., {Lagrange}, A.~M., {Lemoine}, M., \&
  {Vidal-Madjar}, A. 1995, \aap, 304, 461

\bibitem[{{Lallement} {et~al.}(1996){Lallement}, {Linsky}, {Lequeux}, \&
  {Baranov}}]{Lallement1996}
{Lallement}, R., {Linsky}, J.~L., {Lequeux}, J., \& {Baranov}, V.~B. 1996,
  \ssr, 78, 299

\bibitem[{{Lallement} {et~al.}(2014){Lallement}, {Vergely}, {Valette},
  {Puspitarini}, {Eyer}, \& {Casagrande}}]{Lallement2014}
{Lallement}, R., {Vergely}, J.-L., {Valette}, B., {et~al.} 2014, \aap, 561, A91

\bibitem[{{Lallement} {et~al.}(2011){Lallement}, {Welsh}, {Barstow}, \&
  {Casewell}}]{Lallement2011}
{Lallement}, R., {Welsh}, B.~Y., {Barstow}, M.~A., \& {Casewell}, S.~L. 2011,
  \aap, 533, A140

\bibitem[{{Lehner} {et~al.}(2003){Lehner}, {Jenkins}, {Gry}, {Moos}, {Chayer},
  \& {Lacour}}]{Lehner2003}
{Lehner}, N., {Jenkins}, E.~B., {Gry}, C., {et~al.} 2003, \apj, 595, 858

\bibitem[{{Lehner} {et~al.}(2011){Lehner}, {Zech}, {Howk}, \&
  {Savage}}]{Lehner2011}
{Lehner}, N., {Zech}, W.~F., {Howk}, J.~C., \& {Savage}, B.~D. 2011, \apj, 727,
  46

\bibitem[{{Lodders}(2003)}]{Lodders2003}
{Lodders}, K. 2003, \apj, 591, 1220

\bibitem[{{Lodders} \& {Palme}(2009)}]{Lodders.Palme2009}
{Lodders}, K. \& {Palme}, H. 2009, Meteoritics and Planetary Science
  Supplement, 72, 5154

\bibitem[{{Magnani} {et~al.}(1985){Magnani}, {Blitz}, \& {Mundy}}]{Magnani1985}
{Magnani}, L., {Blitz}, L., \& {Mundy}, L. 1985, \apj, 295, 402

\bibitem[{{Malamut} {et~al.}(2014){Malamut}, {Redfield}, {Linsky}, {Wood}, \&
  {Ayres}}]{Malamut2014}
{Malamut}, C., {Redfield}, S., {Linsky}, J.~L., {Wood}, B.~E., \& {Ayres},
  T.~R. 2014, \apj, 787, 75

\bibitem[{{Marr} \& {West}(1976)}]{Marr.West1976}
{Marr}, G.~V. \& {West}, J.~B. 1976, Atomic Data and Nuclear Data Tables, 18,
  497

\bibitem[{{Mathis} {et~al.}(1983){Mathis}, {Mezger}, \& {Panagia}}]{Mathis1983}
{Mathis}, J.~S., {Mezger}, P.~G., \& {Panagia}, N. 1983, \aap, 128, 212

\bibitem[{{McCammon} \& {Sanders}(1990)}]{McCammon.Sanders1990}
{McCammon}, D. \& {Sanders}, W.~T. 1990, \araa, 28, 657

\bibitem[{{McComas} {et~al.}(2012){McComas}, {Alexashov}, {Bzowski}, {Fahr},
  {Heerikhuisen}, {Izmodenov}, {Lee}, {M{\"o}bius}, {Pogorelov}, {Schwadron},
  \& {Zank}}]{McComas2012}
{McComas}, D.~J., {Alexashov}, D., {Bzowski}, M., {et~al.} 2012, Science, 336,
  1291

\bibitem[{{McComas} {et~al.}(2015){McComas}, {Bzowski}, {Fuselier}, {Frisch},
  {Galli}, {Izmodenov}, {Katushkina}, {Kubiak}, {Lee}, {Leonard}, {M{\"o}bius},
  {Park}, {Schwadron}, {Sok{\'o}{\l}}, {Swaczyna}, {Wood}, \&
  {Wurz}}]{McComas2015}
{McComas}, D.~J., {Bzowski}, M., {Fuselier}, S.~A., {et~al.} 2015, \apjs, 220,
  22

\bibitem[{{McKee} \& {Cowie}(1977)}]{McKee.Cowie1977}
{McKee}, C.~F. \& {Cowie}, L.~L. 1977, \apj, 215, 213

\bibitem[{{Meyer} {et~al.}(2006){Meyer}, {Lauroesch}, {Heiles}, {Peek}, \&
  {Engelhorn}}]{Meyer2006}
{Meyer}, D.~M., {Lauroesch}, J.~T., {Heiles}, C., {Peek}, J.~E.~G., \&
  {Engelhorn}, K. 2006, \apjl, 650, L67

\bibitem[{{Meyer} {et~al.}(2012){Meyer}, {Lauroesch}, {Peek}, \&
  {Heiles}}]{Meyer2012}
{Meyer}, D.~M., {Lauroesch}, J.~T., {Peek}, J.~E.~G., \& {Heiles}, C. 2012,
  \apj, 752, 119

\bibitem[{{M{\"o}bius} {et~al.}(2004){M{\"o}bius}, {Bzowski}, {Chalov}, {Fahr},
  {Gloeckler}, {Izmodenov}, {Kallenbach}, {Lallement}, {McMullin}, {Noda},
  {Oka}, {Pauluhn}, {Raymond}, {Ruci{\'n}ski}, {Skoug}, {Terasawa}, {Thompson},
  {Vallerga}, {von Steiger}, \& {Witte}}]{Moebius2004}
{M{\"o}bius}, E., {Bzowski}, M., {Chalov}, S., {et~al.} 2004, \aap, 426, 897

\bibitem[{{Morales} {et~al.}(2001){Morales}, {Orozco}, {G{\'o}mez}, {Trapero},
  {Talavera}, {Bowyer}, {Edelstein}, {Korpela}, {Lampton}, \&
  {Drake}}]{Morales2001}
{Morales}, C., {Orozco}, V., {G{\'o}mez}, J.~F., {et~al.} 2001, \apj, 552, 278

\bibitem[{{Morton}(2003)}]{Morton2003}
{Morton}, D.~C. 2003, \apjs, 149, 205

\bibitem[{{Nussbaumer} \& {Storey}(1981)}]{Nussbaumer.Storey1981}
{Nussbaumer}, H. \& {Storey}, P.~J. 1981, \aap, 96, 91

\bibitem[{{Nussbaumer} \& {Storey}(1983)}]{Nussbaumer.Storey1983}
{Nussbaumer}, H. \& {Storey}, P.~J. 1983, \aap, 126, 75

\bibitem[{{Nussbaumer} \& {Storey}(1986)}]{Nussbaumer.Storey1986}
{Nussbaumer}, H. \& {Storey}, P.~J. 1986, \aaps, 64, 545

\bibitem[{{Oegerle} {et~al.}(2005){Oegerle}, {Jenkins}, {Shelton}, {Bowen}, \&
  {Chayer}}]{Oegerle2005}
{Oegerle}, W.~R., {Jenkins}, E.~B., {Shelton}, R.~L., {Bowen}, D.~V., \&
  {Chayer}, P. 2005, \apj, 622, 377

\bibitem[{{Peek} {et~al.}(2011){Peek}, {Heiles}, {Peek}, {Meyer}, \&
  {Lauroesch}}]{Peek2011}
{Peek}, J.~E.~G., {Heiles}, C., {Peek}, K.~M.~G., {Meyer}, D.~M., \&
  {Lauroesch}, J.~T. 2011, \apj, 735, 129

\bibitem[{{Rafikov} \& {Garmilla}(2012)}]{Rafikov.Garmilla2012}
{Rafikov}, R.~R. \& {Garmilla}, J.~A. 2012, \apj, 760, 123

\bibitem[{{Redfield} \& {Falcon}(2008)}]{Redfield.Falcon2008}
{Redfield}, S. \& {Falcon}, R.~E. 2008, \apj, 683, 207

\bibitem[{{Redfield} \& {Linsky}(2002)}]{Redfield.Linsky2002}
{Redfield}, S. \& {Linsky}, J.~L. 2002, \apjs, 139, 439

\bibitem[{{Redfield} \& {Linsky}(2004{\natexlab{a}})}]{Redfield.Linsky2004a}
{Redfield}, S. \& {Linsky}, J.~L. 2004{\natexlab{a}}, \apj, 602, 776

\bibitem[{{Redfield} \& {Linsky}(2004{\natexlab{b}})}]{Redfield.Linsky2004b}
{Redfield}, S. \& {Linsky}, J.~L. 2004{\natexlab{b}}, \apj, 613, 1004

\bibitem[{{Redfield} \& {Linsky}(2008)}]{Redfield.Linsky2008}
{Redfield}, S. \& {Linsky}, J.~L. 2008, \apj, 673, 283, (RL08)

\bibitem[{{Rogerson} {et~al.}(1973{\natexlab{a}}){Rogerson}, {Spitzer},
  {Drake}, {Dressler}, {Jenkins}, {Morton}, \& {York}}]{Rogerson1973a}
{Rogerson}, J.~B., {Spitzer}, L., {Drake}, J.~F., {et~al.} 1973{\natexlab{a}},
  \apjl, 181, L97

\bibitem[{{Rogerson} {et~al.}(1973{\natexlab{b}}){Rogerson}, {York}, {Drake},
  {Jenkins}, {Morton}, \& {Spitzer}}]{Rogerson1973b}
{Rogerson}, J.~B., {York}, D.~G., {Drake}, J.~F., {et~al.} 1973{\natexlab{b}},
  \apjl, 181, L110

\bibitem[{{Samson} {et~al.}(1994){Samson}, {He}, {Yin}, \&
  {Haddad}}]{Samson1994}
{Samson}, J.~A.~R., {He}, Z.~X., {Yin}, L., \& {Haddad}, G.~N. 1994, Journal of
  Physics B Atomic Molecular Physics, 27, 887

\bibitem[{{Savage} \& {Lehner}(2006)}]{Savage.Lehner2006}
{Savage}, B.~D. \& {Lehner}, N. 2006, \apjs, 162, 134

\bibitem[{{Savage} \& {Sembach}(1996)}]{Savage.Sembach1996}
{Savage}, B.~D. \& {Sembach}, K.~R. 1996, \araa, 34, 279

\bibitem[{{Sembach} {et~al.}(1997){Sembach}, {Savage}, \&
  {Tripp}}]{Sembach1997}
{Sembach}, K.~R., {Savage}, B.~D., \& {Tripp}, T.~M. 1997, \apj, 480, 216

\bibitem[{{Seon} {et~al.}(2011){Seon}, {Edelstein}, {Korpela}, {Witt}, {Min},
  {Han}, {Shinn}, {Kim}, \& {Park}}]{Seon2011}
{Seon}, K.-I., {Edelstein}, J., {Korpela}, E., {et~al.} 2011, \apjs, 196, 15

\bibitem[{{Shull} \& {van Steenberg}(1982)}]{Shull.VanSteenberg1982}
{Shull}, J.~M. \& {van Steenberg}, M. 1982, \apjs, 48, 95

\bibitem[{{Slavin}(1989)}]{Slavin1989}
{Slavin}, J.~D. 1989, \apj, 346, 718

\bibitem[{{Slavin} \& {Frisch}(2006)}]{Slavin.Frisch2006}
{Slavin}, J.~D. \& {Frisch}, P.~C. 2006, \apjl, 651, L37

\bibitem[{{Slavin} \& {Frisch}(2007)}]{Slavin.Frisch2007}
{Slavin}, J.~D. \& {Frisch}, P.~C. 2007, \ssr, 130, 409

\bibitem[{{Slavin} \& {Frisch}(2008)}]{Slavin.Frisch2008}
{Slavin}, J.~D. \& {Frisch}, P.~C. 2008, \aap, 491, 53

\bibitem[{{Slavin} {et~al.}(1993){Slavin}, {Shull}, \& {Begelman}}]{Slavin1993}
{Slavin}, J.~D., {Shull}, J.~M., \& {Begelman}, M.~C. 1993, \apj, 407, 83

\bibitem[{{Snowden} {et~al.}(2014){Snowden}, {Chiao}, {Collier}, {Porter},
  {Thomas}, {Cravens}, {Robertson}, {Galeazzi}, {Uprety}, {Ursino},
  {Koutroumpa}, {Kuntz}, {Lallement}, {Puspitarini}, {Lepri}, {McCammon},
  {Morgan}, \& {Walsh}}]{Snowden2014}
{Snowden}, S.~L., {Chiao}, M., {Collier}, M.~R., {et~al.} 2014, \apjl, 791, L14

\bibitem[{{Snowden} {et~al.}(1997){Snowden}, {Egger}, {Freyberg}, {McCammon},
  {Plucinsky}, {Sanders}, {Schmitt}, {Truemper}, \& {Voges}}]{Snowden1997}
{Snowden}, S.~L., {Egger}, R., {Freyberg}, M.~J., {et~al.} 1997, \apj, 485, 125

\bibitem[{{Snowden} {et~al.}(2015){Snowden}, {Heiles}, {Koutroumpa}, {Kuntz},
  {Lallement}, {McCammon}, \& {Peek}}]{Snowden2015}
{Snowden}, S.~L., {Heiles}, C., {Koutroumpa}, D., {et~al.} 2015, \apj, 806, 119

\bibitem[{{Spitzer}(1978)}]{Spitzer1978}
{Spitzer}, L. 1978, {Physical processes in the interstellar medium} (New York
  Wiley-Interscience)

\bibitem[{{Spitzer}(1996)}]{Spitzer1996}
{Spitzer}, Jr., L. 1996, \apjl, 458, L29

\bibitem[{{Spitzer} \& {Jenkins}(1975)}]{Spitzer.Jenkins1975}
{Spitzer}, Jr., L. \& {Jenkins}, E.~B. 1975, \araa, 13, 133

\bibitem[{{Stancil} {et~al.}(1999){Stancil}, {Schultz}, {Kimura}, {Gu},
  {Hirsch}, \& {Buenker}}]{Stancil1999}
{Stancil}, P.~C., {Schultz}, D.~R., {Kimura}, M., {et~al.} 1999, \aaps, 140,
  225

\bibitem[{{Tayal}(2008)}]{Tayal2008}
{Tayal}, S.~S. 2008, \aap, 486, 629

\bibitem[{{Vallerga}(1998)}]{Vallerga1998}
{Vallerga}, J. 1998, \apj, 497, 921

\bibitem[{{Verner} {et~al.}(1996){Verner}, {Ferland}, {Korista}, \&
  {Yakovlev}}]{Verner1996}
{Verner}, D.~A., {Ferland}, G.~J., {Korista}, K.~T., \& {Yakovlev}, D.~G. 1996,
  \apj, 465, 487

\bibitem[{{Verner} \& {Yakovlev}(1995)}]{Verner.Yakovlev1995}
{Verner}, D.~A. \& {Yakovlev}, D.~G. 1995, \aaps, 109

\bibitem[{{Verschuur}(1969)}]{Verschuur1969}
{Verschuur}, G.~L. 1969, \aplett, 4, 85

\bibitem[{{Verschuur} \& {Knapp}(1971)}]{Verschuur.Knapp1971}
{Verschuur}, G.~L. \& {Knapp}, G.~R. 1971, \aj, 76, 403

\bibitem[{{Vidal-Madjar} \& {Ferlet}(2002)}]{VidalMadjar.Ferlet2002}
{Vidal-Madjar}, A. \& {Ferlet}, R. 2002, \apjl, 571, L169

\bibitem[{{Wakker} {et~al.}(2012){Wakker}, {Savage}, {Fox}, {Benjamin}, \&
  {Shapiro}}]{Wakker2012}
{Wakker}, B.~P., {Savage}, B.~D., {Fox}, A.~J., {Benjamin}, R.~A., \&
  {Shapiro}, P.~R. 2012, \apj, 749, 157

\bibitem[{{Wang} {et~al.}(2010){Wang}, {Wan}, \& {Zhou}}]{Wang2010}
{Wang}, G., {Wan}, J., \& {Zhou}, X. 2010, Journal of Physics B Atomic
  Molecular Physics, 43, 035001

\bibitem[{{Wehlitz} {et~al.}(2007){Wehlitz}, {Lukic}, \&
  {Juranic}}]{Wehlitz2007}
{Wehlitz}, R., {Lukic}, D., \& {Juranic}, P.~N. 2007, Journal of Physics B
  Atomic Molecular Physics, 40, 2385

\bibitem[{{Weingartner} \& {Draine}(2001)}]{Weingartner.Draine2001}
{Weingartner}, J.~C. \& {Draine}, B.~T. 2001, \apj, 563, 842

\bibitem[{{Welsh} {et~al.}(2010){Welsh}, {Lallement}, {Vergely}, \&
  {Raimond}}]{Welsh2010}
{Welsh}, B.~Y., {Lallement}, R., {Vergely}, J.-L., \& {Raimond}, S. 2010, \aap,
  510, A54

\bibitem[{{Williamson} {et~al.}(1974){Williamson}, {Sanders}, {Kraushaar},
  {McCammon}, {Borken}, \& {Bunner}}]{Williamson1974}
{Williamson}, F.~O., {Sanders}, W.~T., {Kraushaar}, W.~L., {et~al.} 1974,
  \apjl, 193, L133

\bibitem[{{Wolfire} {et~al.}(2003){Wolfire}, {McKee}, {Hollenbach}, \&
  {Tielens}}]{Wolfire2003}
{Wolfire}, M.~G., {McKee}, C.~F., {Hollenbach}, D., \& {Tielens}, A.~G.~G.~M.
  2003, \apj, 587, 278

\bibitem[{{Wood} {et~al.}(2005){Wood}, {Redfield}, {Linsky}, {M{\"u}ller}, \&
  {Zank}}]{Wood2005}
{Wood}, B.~E., {Redfield}, S., {Linsky}, J.~L., {M{\"u}ller}, H.-R., \& {Zank},
  G.~P. 2005, \apjs, 159, 118

\bibitem[{{Wood} {et~al.}(2002){Wood}, {Redfield}, {Linsky}, \&
  {Sahu}}]{Wood2002}
{Wood}, B.~E., {Redfield}, S., {Linsky}, J.~L., \& {Sahu}, M.~S. 2002, \apj,
  581, 1168

\bibitem[{{Zirnstein} {et~al.}(2016){Zirnstein}, {Heerikhuisen}, {Funsten},
  {Livadiotis}, {McComas}, \& {Pogorelov}}]{Zirnstein2016}
{Zirnstein}, E.~J., {Heerikhuisen}, J., {Funsten}, H.~O., {et~al.} 2016, \apjl,
  818, L18

\bibitem[{{Zsarg{\'o}} {et~al.}(2003){Zsarg{\'o}}, {Sembach}, {Howk}, \&
  {Savage}}]{Zsargo2003}
{Zsarg{\'o}}, J., {Sembach}, K.~R., {Howk}, J.~C., \& {Savage}, B.~D. 2003,
  \apj, 586, 1019

\end{thebibliography}
\bibpunct{(}{)}{;}{a}{}{,} 

\end{document}